\documentclass[structabstract,longauth]{aa}
\usepackage{graphicx}
\usepackage{lscape}
\usepackage{amsmath}
\usepackage{amssymb}
\usepackage{epsfig}
\usepackage{hyperref}
\usepackage{color}

\usepackage{txfonts}
\usepackage{longtable,lscape}
\usepackage{natbib}
\usepackage{longtable}
\usepackage{here}
\bibpunct{(}{)}{;}{a}{}{,} 

\newcommand{\corot}{\textsl{CoRoT}}  

\def\ms{\,m\,s$^{-1}$}               
\def\kms{\,km\,s$^{-1}$}             
\def\teff{$T_{\rm eff}$}              
\def\logg{$log\,g$}                   
\def\vsini{$v$\,sin\,$i$}             
\def\ms{\hbox{\,m\,s$^{-1}$}}         
\def\m2s2{\hbox{\,m$^{2}$\,s$^{-2}$}} 
\def\kms{\hbox{\,km\,s$^{-1}$}}       
\def\gcm3{\hbox{\,g\,cm$^{-3}$}}      
\def\vsini{\hbox{$v$\,sin\,$i$}}      

\def\degr{\hbox{$^\circ$}}

\begin{document}
\title{Planetary transit candidates in the CoRoT LRa01 field\thanks{The CoRoT 
space mission, launched on December 27th 2006, has been developed and is operated 
by CNES, with the contribution of Austria, Belgium, Brazil , ESA (RSSD and Science 
Programme), Germany and Spain.}}

\author{L.~Carone\inst{\ref{Koeln}}
\and D.~Gandolfi\inst{\ref{ESA},\ref{Tautenburg}}
\and J.~Cabrera\inst{\ref{DLR},\ref{LUTh}}
\and A.~P.~Hatzes\inst{\ref{Tautenburg}}
\and H.~J.~Deeg\inst{\ref{IAC},\ref{La Laguna}}
\and Sz.~Csizmadia\inst{\ref{DLR}}
\and M.~P\"atzold\inst{\ref{Koeln}}
\and J.~Weingrill\inst{\ref{Graz}}
\and S.~Aigrain\inst{\ref{Oxford}}
\and R.~Alonso\inst{\ref{Geneve}}
\and A.~Alapini\inst{\ref{Exeter}}
\and J.-M.~Almenara\inst{\ref{IAC},\ref{La Laguna},\ref{LAM}}
\and M.~Auvergne\inst{\ref{LESIA}}
\and A.~Baglin\inst{\ref{LESIA}}
\and P.~Barge\inst{\ref{LAM}}
\and A.~S.~Bonomo\inst{\ref{LAM}}
\and P.~Bord\'e\inst{\ref{IAS}}
\and F.~Bouchy\inst{\ref{IAP},\ref{OHP}}
\and H.~Bruntt\inst{\ref{LESIA}}
\and S.~Carpano\inst{\ref{ESA}}
\and W.~D.~Cochran \inst{\ref{McDonald}}
\and M.~Deleuil\inst{\ref{LAM}}
\and R.~F.~D\'iaz\inst{\ref{IAP},\ref{OHP}}
\and S.~Dreizler\inst{\ref{Goettingen}}
\and R.~Dvorak\inst{\ref{Wien}}
\and J.~Eisl\"offel\inst{\ref{Tautenburg}}
\and P.~Eigm\"uller\inst{\ref{Tautenburg}}
\and M.~Endl\inst{\ref{McDonald}}
\and A.~Erikson\inst{\ref{DLR}}
\and S.~Ferraz-Mello\inst{\ref{IAG}}
\and M.~Fridlund\inst{\ref{ESA}}
\and J.-C.~Gazzano\inst{\ref{LAM},\ref{OCA}}
\and N.~Gibson\inst{\ref{Oxford},\ref{Exeter}}
\and M.~Gillon\inst{\ref{Geneve},\ref{Liege}}
\and P.~Gondoin\inst{\ref{ESA}}
\and S.~Grziwa\inst{\ref{Koeln}}
\and E.W.~G\"unther\inst{\ref{Tautenburg}}
\and T.~Guillot\inst{\ref{OCA}}
\and M.~Hartmann\inst{\ref{Tautenburg}}
\and M.~Havel\inst{\ref{OCA}}
\and G.~H\'ebrard\inst{\ref{IAP}}
\and L.~Jorda\inst{\ref{LAM}}
\and P. Kabath\inst{\ref{DLR},\ref{ESO-Chile}}
\and A.~L\'eger\inst{\ref{IAS}}
\and A.~Llebaria\inst{\ref{LAM}}
\and H.~Lammer\inst{\ref{Graz}}
\and C.~Lovis\inst{\ref{Geneve}}
\and P.~J.~MacQueen\inst{\ref{McDonald}}
\and M.~Mayor\inst{\ref{Geneve}}
\and T.~Mazeh\inst{\ref{Tel Aviv}}
\and C.~Moutou\inst{\ref{LAM}}
\and L.~Nortmann\inst{\ref{Goettingen}}
\and A.~Ofir\inst{\ref{Tel Aviv}}
\and M.~Ollivier\inst{\ref{IAS}}
\and H. Parviainen\inst{\ref{IAC},\ref{La Laguna}}
\and F.~Pepe\inst{\ref{Geneve}}
\and F.~Pont\inst{\ref{Exeter}}
\and D.~Queloz\inst{\ref{Geneve}}
\and M.~Rabus\inst{\ref{IAC},\ref{La Laguna},\ref{Chile}}
\and H.~Rauer\inst{\ref{DLR},\ref{ZAA}}
\and C.~R\'egulo\inst{\ref{IAC},\ref{La Laguna}}
\and S.~Renner\inst{\ref{DLR},\ref{Lille},\ref{IMCCE}}
\and R.~de la Reza\inst{\ref{ONRdJ}}
\and D.~Rouan\inst{\ref{LESIA}}
\and A.~Santerne\inst{\ref{LAM}}
\and B.~Samuel\inst{\ref{IAS}}
\and J.~Schneider\inst{\ref{LUTh}}
\and A.~Shporer\inst{\ref{Tel Aviv},\ref{Las_Cumbres}}
\and B.~Stecklum\inst{\ref{Tautenburg}}
\and L.~Tal-Or\inst{\ref{Tel Aviv}}
\and B.~Tingley\inst{\ref{IAC},\ref{La Laguna}}
\and S.~Udry \inst{\ref{Geneve}}
\and G.~Wuchterl\inst{\ref{Tautenburg}}
}

\institute{
     Rheinisches Institut f\"ur Umweltforschung, Abteilung Planetenforschung, an der Universit\"at zu K\"oln, Aachener Strasse 209, 50931, Germany\\ \email{lcarone@uni-koeln.de}\label{Koeln}
\and Research and Scientific Support Department, ESTEC/ESA, PO Box 299, 2200 AG Noordwijk, The Netherlands\\ \email{davide.gandolfi@esa.int}\label{ESA}
\and Th\"uringer Landessternwarte, Sternwarte 5, Tautenburg, D-07778 Tautenburg, Germany\label{Tautenburg}
\and Institute of Planetary Research, German Aerospace Center, Rutherfordstrasse 2, 12489 Berlin, Germany\label{DLR}
\and LUTH, Observatoire de Paris, CNRS, Universit\'e Paris Diderot; 5 place Jules Janssen, 92195 Meudon, France\label{LUTh}
\and Instituto de Astrof{\'i}sica de Canarias, E-38205 La Laguna, Tenerife, Spain\label{IAC}
\and Departamento de Astrof\'isica, Universidad de La Laguna, 38206 La Laguna, Tenerife, Spain\label{La Laguna}
\and Space Research Institute, Austrian Academy of Science, Schmiedlstr. 6, A-8042 Graz, Austria\label{Graz}
\and Oxford Astrophyiscs, Denys Wilkinson Building, Keble Road, Oxford OX1 3RH\label{Oxford}
\and Observatoire de l'Universit\'e de Gen\`eve, 51 chemin des Maillettes, 1290 Sauverny, Switzerland\label{Geneve}
\and School of Physics, University of Exeter, Stocker Road, Exeter EX4 4QL, United Kingdom\label{Exeter}
\and Laboratoire d'Astrophysique de Marseille, CNRS \& University of Provence, 38 rue Fr\'ed\'eric Joliot-Curie, 13388 Marseille cedex 13, France\label{LAM}
\and LESIA, Observatoire de Paris, Place Jules Janssen, 92195 Meudon cedex, France\label{LESIA}
\and Institut d'Astrophysique Spatiale, Universit\'e Paris XI, F-91405 Orsay, France\label{IAS}
\and Institut d'Astrophysique de Paris, UMR7095 CNRS, Universit\'e Pierre \& Marie Curie, 98bis boulevard Arago, 75014 Paris, France\label{IAP}
\and Observatoire de Haute Provence, 04670 Saint Michel l'Observatoire, France\label{OHP}
\and McDonald Observatory, University of Texas at Austin, Austin, TX 78712, USA\label{McDonald}
\and Institut f\"ur Astrophysik, Georg-August-Universit\"at, Friedrich-Hund-Platz 1, D-37077 G\"ottingen, Germany\label{Goettingen}
\and University of Vienna, Institute of Astronomy, T\"urkenschanzstrasse 17, A-1180 Vienna, Austria\label{Wien}
\and IAG, University of S\~ao Paulo, Brasil\label{IAG}
\and Universit\'e de Nice-Sophia Antipolis, CNRS UMR 6202, Observatoire de la C\^ote d'Azur, BP 4229, 06304 Nice Cedex 4, France\label{OCA}
\and University of Li\`ege, All\'ee du 6 ao\^ut 17, Sart Tilman, Li\`ege 1, Belgium\label{Liege}
\and European Southern Observatory, Alonso de C\'ordova 3107, Casilla 19001, Santiago de Chile, Chile\label{ESO-Chile}
\and School of Physics and Astronomy, Raymond and Beverly Sackler Faculty of Exact Sciences, Tel Aviv University, Tel Aviv, Israel\label{Tel Aviv}
\and Departamento de Astronom\'ia y Astrof\'sica, Pontificia Universidad Cat\'olica de Chile, Casilla 306, Santiago 22, Chile\label{Chile}
\and Center for Astronomy and Astrophysics, TU Berlin, Hardenbergstr. 36, 10623 Berlin, Germany\label{ZAA}
\and Laboratoire d'Astronomie de Lille, Universit\'e de Lille 1, 1 impasse de l'Observatoire, 59000 Lille, France\label{Lille}
\and Institut de M\'ecanique C\'eleste et de Calcul des Eph\'em\'erides, UMR 8028 du CNRS, 77 avenue Denfert-Rochereau, 75014 Paris, France\label{IMCCE}
\and Observat\'orio Nacional, Rio de Janeiro, Brazil\label{ONRdJ}
\and Las Cumbres Observatory Global Telescope Network, Inc., 6740 Cortona Drive, Suite 102, Santa Barbara, California 93117, USA\label{Las_Cumbres}
}
   \date{Received ...; accepted ...}


  \abstract
   {\corot\ is a pioneering space mission whose primary goals are stellar
    seismology and extrasolar planets search. Its surveys of large stellar 
    fields generate numerous planetary candidates whose lightcurves have 
    transit-like features. An extensive analytical and observational 
    follow-up effort is undertaken to classify these candidates.}
   {The list of planetary transit candidates from the \corot\ LRa01 star field 
   in the Monoceros constellation towards the Galactic anti-center is presented. 
   The \corot\ observations of LRa01 lasted from 24 October 2007 to 3 March 2008.}
   {7\,470 chromatic and 3\,938 monochromatic lightcurves were acquired and analysed. 
   Instrumental noise and stellar variability were treated with several filtering tools 
   by different teams from the \corot\ community. Different transit search algorithms 
   were applied to the lightcurves.}
   {Fifty-one stars were classified as planetary transit candidates in LRa01. Thirty-seven 
   (i.e., $73$\,\% of all candidates) are ``good'' planetary candidates based on photometric 
   analysis only. Thirty-two (i.e., 87\,\% of the ``good" candidates) have been followed-up.
   At the time of this writing twenty-two cases have been solved and five planets have been
   discovered: three transiting hot-Jupiters (\corot-5b, \corot-12b, and \corot-21b), the 
   first terrestrial transiting planet (\corot-7b), and another planet in the same system 
   (\corot-7c, detected by radial velocity survey only). Evidences of another non-transiting 
   planet in the \corot-7 system, namely \corot-7d, have been recently found.}
   {}


\titlerunning{Planetary transit candidates in the CoRoT LRa01 field}
\authorrunning{Carone et al.}

   \maketitle
%

\section{Introduction}
\label{sec: intro}

This paper summarizes the planetary candidates found in the LRa01 
exoplanet star field and some preliminary scientific results from 
the combination of \corot\ photometry with ground based follow-up 
observations. The \corot\ IRa01 and LRc01 runs have already been 
reported by \citet{Carpano2009} and \citet{Moutou2009}, and 
\citet{Cabrera2009}, respectively.

The LRa01 run, from 24 October 2007 to 3 March 2008, was the second 
long pointing of \corot\ after the LRc01 field \citep{Cabrera2009}. 
The LRa01 star field contains 11\,408 pre-selected stars covering a 
sky-area within the coordinates $06^h45^m11.2^s\le RA\le06^h45^m59.4^s$ 
and $-01\degr27\arcmin21\arcsec\le\delta\le+01\degr6\arcmin23\arcsec$ 
(J2000) in the Monoceros constellation, towards Galactic anti-center 
direction.

Fifty-one transit candidates have been identified in LRa01 
(Tables\,\ref{Tab:all_planets} and \ref{Tab:all_planets_cont}). Four 
transiting planets have been discovered and confirmed: \corot-5b 
\citep{Rauer2009}, \corot-12b \citep{Gillon2010}, and CoRoT-21b 
\citep{Patzold2011}, three Jupiter-size planets with $M_p=0.47$, 
$0.92$, and $\approx 2$\,$M_{Jup}$, respectively; \corot-7b, the 
first terrestrial transiting planet \citep{Leger2009, Queloz2009}. 
A non-transiting planet with a mass of about 8\,$M_{Earth}$ in the 
\corot-7-system, namely \corot-7c, was also detected by radial velocity 
(RV) observations only \citep{Queloz2009}. The potential discovery of 
a third planet in the \corot-7-system, \corot-7d, was reported by 
\citet{Hatzes2010}. A list of photometrically identified binary 
systems is presented in the Table\,\ref{Tab:Binaries}. Identified 
variable stars of the first four \corot\ exoplanet star fields are 
reported in \citet{Debosscher2009}.

The present work reports on the characteristics of the LRa01 star-field 
(Section~\ref{sec: field}), the \corot\ photometry and nature of 
different instrumental systematic effects (Section~\ref{sec:Data}), 
the transit detection (Sections~\ref{sec: detection}), and the observing 
strategy of ground-based follow-ups (Section~\ref{sec: FU}). The 
process of resolving the nature of \corot\ candidates is described in 
\citet{Moutou2009}. A description of all the detected transit candidates 
is presented in Section~\ref{sec: candidates}. Results are discussed 
in terms of detection efficiency compared to previous \corot\ runs 
(Section~\ref{sec: discussion}). At the end of the manuscript, a summary 
is also reported (Section~\ref{sec: summary}).

\section{Field characterisation}
\label{sec: field}

\begin{figure}[t]
  \begin{center}
    \includegraphics[%
      width=0.9\linewidth,%
      height=0.5\textheight,%
      keepaspectratio]{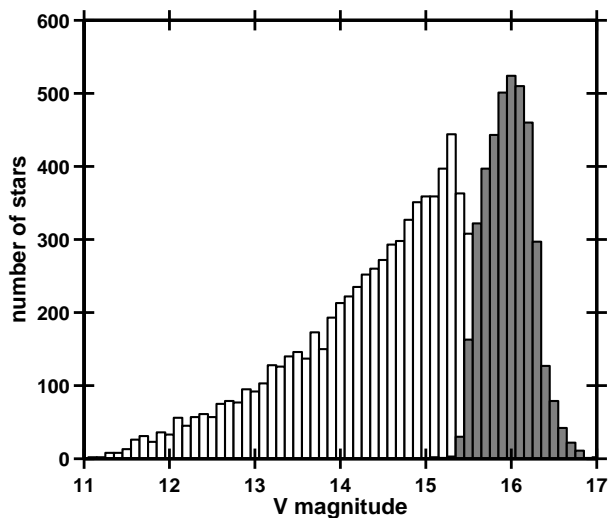}
  \end{center}
  \caption{Histogram of the the visual magnitudes of the stars observed 
           by \corot\ in the LRa01 exoplanet star field. Gray: monochromatic 
           lightcurves. White: chromatic lightcurves. The majority of 
           the targets are relatively faint stars with $V>14$\,mag.}
  \label{fig:m_V}
\end{figure}

During the \corot\ mission preparatory phase, a massive and deep Harris 
$BV$ and Sloan $r'i'$ photometric survey was performed in the \corot\ 
exoplanet fields using the Wide Field Camera (WFC) at the Isaac Newton 
Telescope (INT). The goals were i) to perform a first-order spectral 
classification of the stars in the fields, ii) to determine their 
position with sufficient accuracy for a precise placement of the 
\corot\ photometric masks, and iii) to assess the level of contamination 
from background/foreground objects within a few arc-seconds from the 
\corot\ target stars. The relevant information is collected in the 
Exo-Dat database\footnote{http://lamwws.oamp.fr/exodat/} 
\citep{Deleuil2006,Deleuil2009,Meunier2007}.

\corot\ was designed to fulfil two main objectives: conducting stellar seismology 
studies of interesting stars and searching for extrasolar planets. Astroseismology 
requires high signal-to-noise (S/N) ratio photometry and it is thus focused mainly 
on the study of relatively bright targets: typically $\sim$10 stars with $V$$<$9.5~mag 
are observed in each \corot\ \emph{seismo}-field. On the other hand, the transiting 
exoplanet search requires a large number of targets because of the low probability to 
find planets whose orbits are oriented such that transits can be observed in front of 
their host stars (the probability is about 5\,\% for semi major axes of $0.1$\,AU). 
The selection of the observed \corot\ \emph{seismo}- and \emph{exo}-fields thus 
represent a compromise between these two requirements.

\begin{figure}[t]
\centering
\begin{minipage}[b]{0.75\linewidth}
    \includegraphics[%
      width=0.9\linewidth,%
      keepaspectratio]{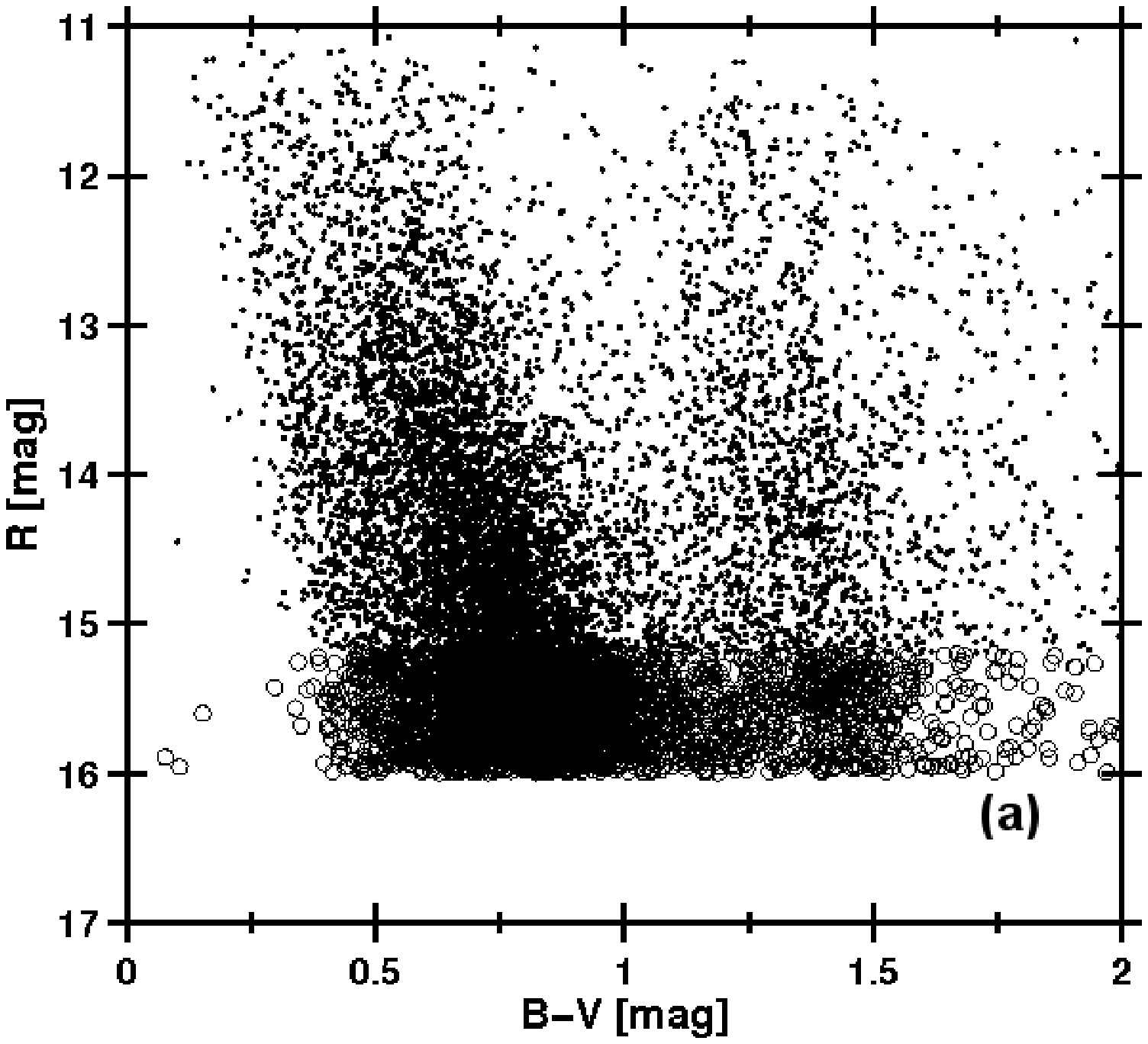}
\end{minipage}

\begin{minipage}[b]{0.75\linewidth}
    \includegraphics[%
      width=0.9\linewidth,%
      keepaspectratio]{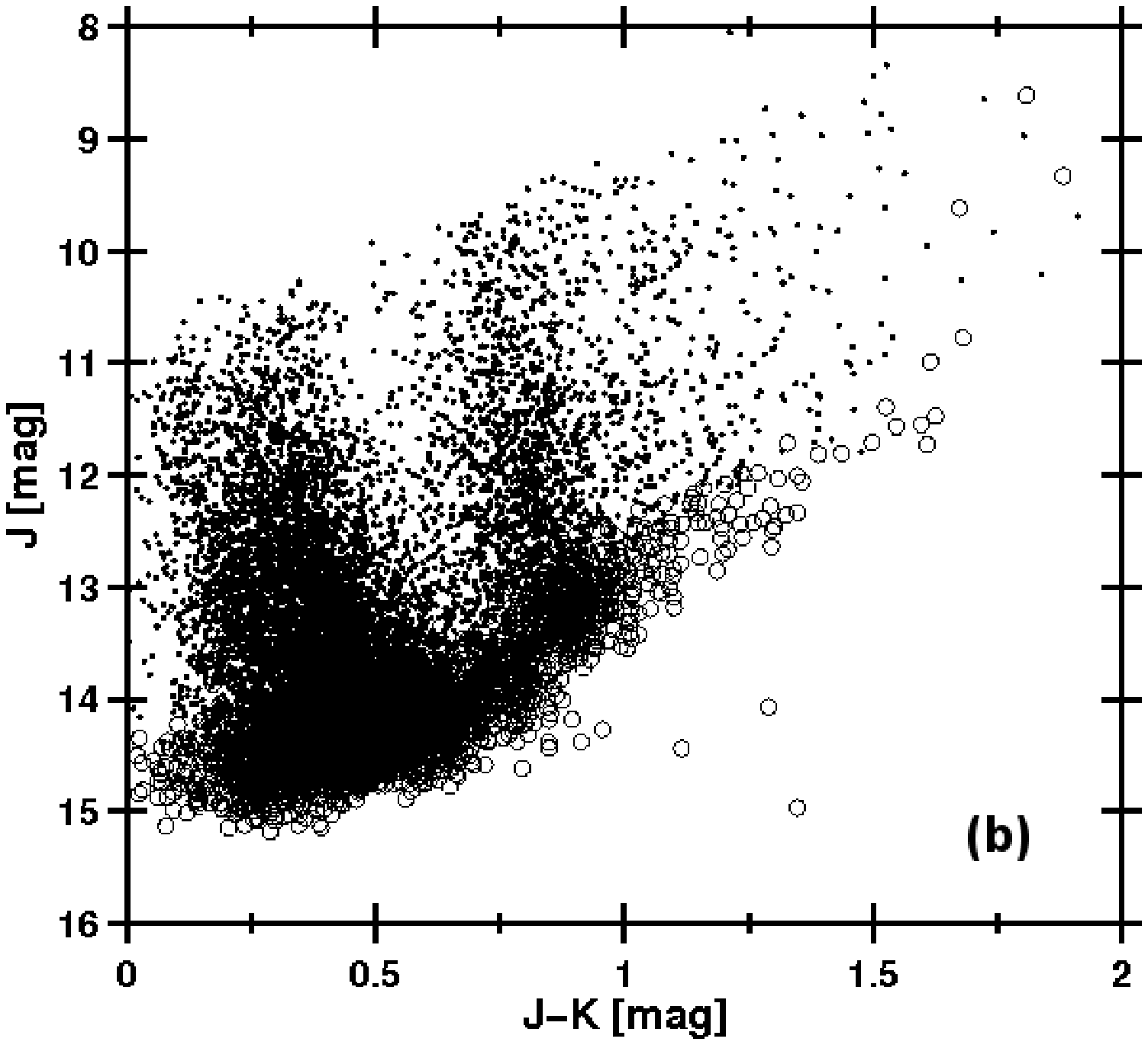}
\end{minipage}
  \caption{\emph{Top panel}: $r^\prime$ versus $B-V$ color-magnitude diagram 
           of the stars in the \corot\ LRa01 star field. The dots mark the 
           bright stars ($r^\prime\lesssim15.2$\,mag), for which three-color 
           \corot\ lightcurves are available. The open circles mark the 
           fainter stars ($r^\prime\gtrsim15.2$\,mag) with monochromatic 
           lightcurves only. Main sequence stars cluster in the left part of 
           the diagram, giant stars in the right part. \emph{Bottom panel}: 
           2MASS $J$ versus $J-Ks$ color-magnitude diagram of the stars in 
           the \corot\ LRa01 field. Again, main sequence stars cluster in 
           the left part of the diagram, giant stars in the right part.}
  \label{fig:giant_dwarfs_color}
\end{figure}

\begin{figure}[t]
  \begin{center}
    \includegraphics[%
      width=0.9\linewidth,%
      height=0.5\textheight,%
      keepaspectratio]{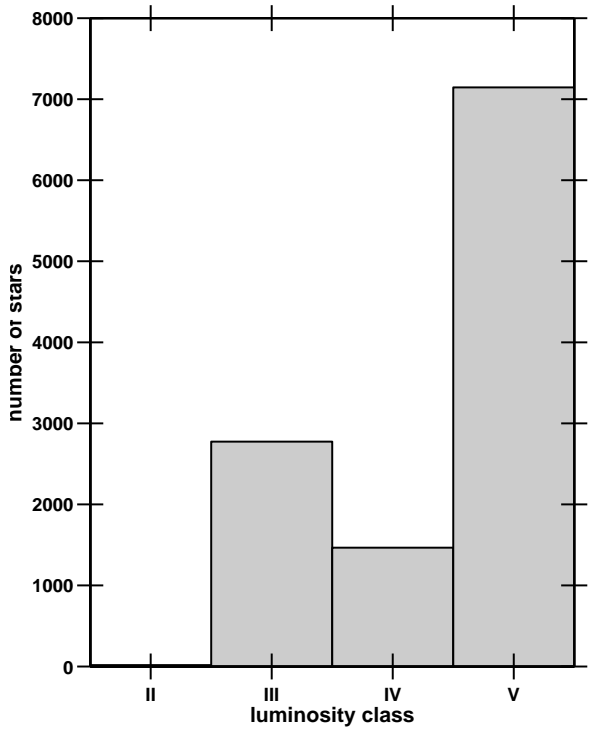}
  \end{center}
  \caption{Histogram of the luminosity classes of the LRa01 star field as 
           derived from Exo-Dat \citep{Deleuil2009}. The majority of the target 
           stars ($\sim62$\,\%) are main sequence dwarf stars (luminosity class V). 
           Sub-giant (luminosity class IV), giant (luminosity class III), 
           and bright giant stars (luminosity class II) make up $\sim$13, $\sim$24, 
           $<$0.2\,\% of the total, respectively. Super-giant stars 
           (luminosity class I; not reported in the histogram) make up 
           $<$0.05\,\%.}
  \label{fig:lumclass}
\end{figure}

11\,408 stars with visual magnitudes $11\lesssim~V\lesssim~17$\,mag were observed 
for the transit search in the LRa01 field (Figure\,\ref{fig:m_V}).
Dwarf stars are optimal targets for the photometric search of extrasolar 
planets \citep{Michel2008,Gondoin2009,Hekker2009}. Therefore, the 
percentage of Sun-like stars in a field is important to estimate the 
detection efficiency. Combining the Exo-Dat optical photometry 
\citep{Deleuil2009} with the near-infrared Two Micron All Sky Survey 
(2MASS) Point Source catalog\footnote{The near-infrared $JHKs$ 2MASS 
catalogue is available at http://irsa.ipac.caltech.edu/applications/2MASS/IM/interactive.html.} 
\citep{Cutri2003} and using color-magnitude diagrams 
(Figure~\ref{fig:giant_dwarfs_color}), the percentage of giant and dwarf 
stars in LRa01 is estimated. Although it is difficult to find a clear 
distinction between these two populations using broad-band photometry
(in particular for stars with $V\gtrsim15$\,mag), 75\,\% of the stars 
observed by \corot\ in LRa01 seem to be dwarf stars 
(Figure~\ref{fig:giant_dwarfs_color}). The distribution of luminosity 
classes derived from Exo-Dat only shows a similar picture 
(Figure~\ref{fig:lumclass}): less than 0.05\,\% of the 
stars in the LRa01 star field are super-giants, less than 0.2\,\% are 
bright giants, $\sim$24\,\% are giants, $\sim$13\,\% are sub-giants, and 
$\sim$62\,\% are dwarf stars. This is of advantage for the search of 
extrasolar planets compared to the LRc01 star field, where $\sim$58\,\% 
of the stars are giant stars \citep{Cabrera2009}. However, this 
analysis is valid on a statistical point of view.  Photometric 
criteria can indeed lead to misclassification of individual stars 
\citep[see][and reference therein]{Klement2011}. The Exo-Dat spectral 
classification, based only on broad-band photometry, suffers some 
uncertainties, the main ones being the star reddening, the unknown chemical 
abundances, and the potential binarity which can result in a wrong 
identification of the spectral and luminosity classes of the stars. Based 
on multi-object, intermediate-resolution spectroscopy performed with 
FLAMES@VLT on a sub-set of stars in LRa01, \citet{Gazzano2010} found that 
the photometrically classified dwarf content is underestimated by about 
15\,\%.

\section{\corot\ photometry, data reduction, and systematic effect}
\label{sec:Data}

\corot\ data are made public after a proprietary period of one year. 
The data from LRa01 were released to the CoIs on 29 October 2008 and 
to the public on 29 October 2009. \corot\ lightcurves are identified 
by either the \corot\ ID, a unique 10 digit number, or the so-called 
\corot\ ``Win ID'' (i.e., window ID). The ``Win ID'' contains the 
identifier E1 or E2, which identifies the CCD 1 or 2 of the exoplanet 
channel, respectively, followed by a 4-digit number. This number 
represents the assigned \corot\ mask. The ``Win ID'' is re-used for 
every run. To identify an individual lightcurve by the ``Win ID'', 
the data acquisition run is required. In addition, a three character 
abbreviation, MON or CHR, is given for the identification of chromatic 
or monochromatic lightcurves, respectively. For example, the 
lightcurve of the star \corot-7, observed during LRa01 and harbouring 
the first transiting terrestrial planet \citep{Leger2009}, is labeled 
by the following identification: LRa01~E2~0165~-~CHR~-~0102708694.

A bi-prism was installed in the exoplanet channel to disperse the flux 
of the observed stars. Three-color photometry is obtained by splitting the 
point-spread function into three sub-areas based on the dispersion 
property of the bi-prism (blue light is stronger dispersed than red light) 
for the targets with visual magnitudes $V \lesssim 15.2$\,mag, i.e., 
about 65\,\% of the lightcurves in this run. Flux from these areas is 
defined as red, green, and blue. The \corot\ color channels, however, do 
not correspond to any standard photometric systems \citep{Auvergne2009}.

\begin{figure}[t]
  \begin{center}
    \includegraphics[%
      width=0.9\linewidth,%
      height=0.5\textheight,%
      keepaspectratio]{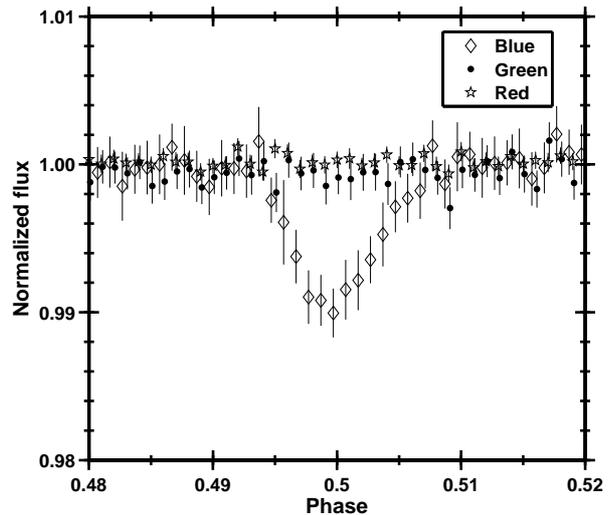}
  \end{center}
  \caption{Phase-plot of the candidate LRa01\,E2\,2597 (\corot\ ID 0102672065) 
           folded at the period $P=8.90$~days in the three color channels red 
           (star), green (circle), and blue (diamond). The lightcurves are 
           normalised to the fluxes in the respective colors. The event is only 
           visible in the blue channel. This indicates a contamination from a 
           background binary system.}
  \label{fig:BEB}
\end{figure}

The chromatic information is helpful to distinguish between achromatic 
planetary transits and chromatic eclipsing binaries. The chromatic 
information is also used to identify false alarms from diluted background 
binaries. After an accurate study of the light contamination inside the 
photometric mask, candidates with 3 sigma significant depth differences in 
the three \corot\ channels are usually flagged as potential contaminating 
eclipsing binaries. As an example, the \corot\ target LRa01\,E2\,2597 has a 
deep transit in the blue channel\footnote{1\,\% deep when normalized to the 
blue flux only, 0.14\,\% deep when normalized to the total flux of all three 
channels.} which is not detected, neither in the green channel nor in the red 
channel, at a 12\,$\sigma$ and 25\,$\sigma$ significance, respectively 
(Figure~\ref{fig:BEB}). If the signal were on target the respective 
transit depths in the red and green channel would have been clearly 
detected. As a consequence, this candidate is identified as a 
contaminating eclipsing binary (CEB). In other cases, i.e., 
LRa01\,E1\,2101 (Section~\ref{LRa01 E1 2101}) and LRa01\,E2\,3156 
(Section~\ref{LRa01 E2 3156}), the signal is detected in only one 
channel but the expected signal strengths in the other channels is 
below the noise threshold. Therefore, we could not exclude such signals 
as arising from a contaminanting eclipsing binary. Indeed, ground-based 
photometric follow-up concluded that the transit signal of LRa01\,E1\,2101 
and LRa01\,E2\,3156 are likely on target.

Several systematic effects need to be filtered out during the data 
reduction to achieve maximum accuracy. The main perturbating factors 
are eclipses (when the spacecraft enters the shadow of the Earth) 
which cause short-term temperature fluctuations on the spacecraft, 
the Earth's gravity and magnetic field, solar and terrestrial infrared 
emissivity, the Earth's albedo, and objects in low-Earth orbit. Known 
instrumental effects like spacecraft jitter are already removed in the 
processing. A detailed description of lightcurve perturbations and 
employed corrections is given in \citet{Auvergne2009}, and 
\citet{Drummond2008} and \citet{Pinheiro2008}, respectively. Details 
about on board data-reduction can be found in \citet{Llebaria2006}.

Irradiation excites single pixels, which are difficult to correct. 
These ``hot-pixels'' appear predominantly in orbit when crossing 
the South Atlantic Anomaly (SAA). A value in the header of each 
lightcurve informs the user how many ``hot-pixels'' were identified 
in the lightcurve. Not all ``hot-pixels'' are identified during the 
data processing. Unfortunately, these events may mimic a transit-like 
signal. If only one pixel is affected, this can be identified by 
comparing the flux in the different color channels (if available).

\begin{figure}[t]
  \begin{center}
    \includegraphics[%
      width=1.0\linewidth,%
      keepaspectratio]{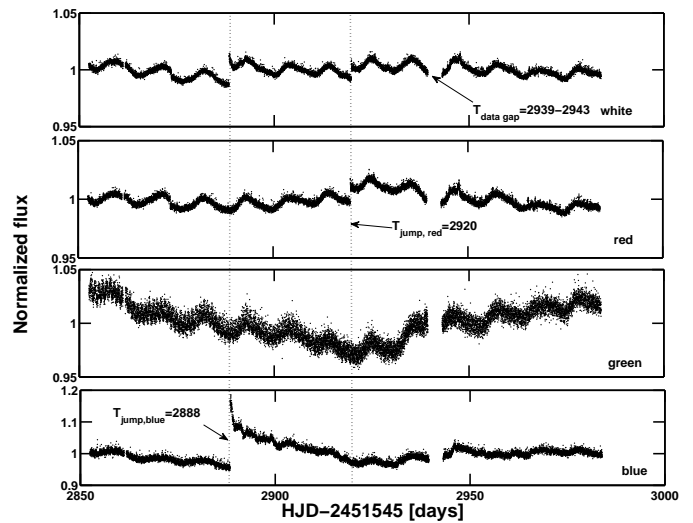}
  \end{center}
  \caption{Chromatic lightcurves of the star LRa01\,E1\,2698. From top to 
           bottom: white (red + green + blue), red, green, and blue 
           lightcurve versus observing time in HJD. The flux of each 
           lightcurve is normalized to 1. The red channel contributes 
           the most to the combined white lightcurve (79\,\% of the flux). 
           10\,\% of the combined white flux originates from the green 
           channel, whereas the blue channel contributes 11\,\% of the 
           total white flux. The dashed lines mark the position of two 
           small ``hot-pixels'' in the blue and red channels that can 
           also be found in the white lightcurve.}
  \label{fig:chr_sys}
\end{figure}

Although the processing pipeline reduces significantly the noise and 
removes the systematics, some instrumental effects still remain. As an 
example, the three color-channel lightcurves of the star LRa01~E1~2698 
(\corot\ ID 0102566329), are plotted in Figure~\ref{fig:chr_sys}. 
Instrumental signatures significantly perturb the lightcurve. On 
2454433 HJD ($T_{jump,blue}=2888$~days on 
Figure~\ref{fig:chr_sys}) a strong ``hot-pixel'' appeared in the blue 
channel. It perturbed the white lightcurve only weakly since the flux 
contribution from the blue channel was small. On 2454465 HJD 
($T_{jump,red}=2920$~days on Figure~\ref{fig:chr_sys}) a 
small ``hot-pixel'' appeared in the red channel. The effect is seen in 
the white lightcurve with almost the same amplitude because the red 
channel contributes most to the overall flux. It should also be noted 
that the relaxation time for this ``hot-pixel'' is very short compared 
to the strong event in the blue channel. The flux after the incident 
settles on a slightly higher level than before. This example proves 
the usefulness of chromatic data to identify ``hot-pixels''.

There is a data gap of 3.68 days between HJD 2454484 and 2454488  
($T_{data\,gap}=2939-2943$~days on Figure~\ref{fig:chr_sys}). A proton 
impact led to a reset of the Data Processing Unit (DPU) 1, which 
is responsible for data collection on the E1 CCD, on 18 January 2008 
at 22:45:57 UT during the SAA crossing. The recording resumed on 22 
January 2008 at 14:37:41 UT. All lightcurves in LRa01 originating from 
CCD 1 contain this data gap.

In order to correct for the effects not removed by the \corot\ pipeline, 
each detection team inside the \corot\ exoplanet team uses a set of 
additional filters before applying the transit search algorithms. A 
description of the various methods applied by each detection team 
to analyse the \corot\ lightcurves is given in \citet{Alapini2008,
Borde2007,Carpano2008,Moutou2005,Moutou2007,Regulo2007,Renner2008,
Grziwa2011}.

\section{Detection of \corot\ candidates}
\label{sec: detection}

The transit candidate selection for each run is performed twice: by the 
``Alarm Mode'', during the on-board observation, and by the \corot\ detection 
teams, after completion of the on-board observation and the full reduction 
of the \corot\ data. 

The ``Alarm Mode'' checks for planetary transits every two weeks, when 
the \corot\ observations are still ongoing \citep{Surace2008}. If the 
``Alarm Mode'' detects a promising candidate with a planetary transit-like 
event, the sampling for this specific lightcurve is increased from 512 to 
32 seconds. A preliminary candidate list is created and the candidate 
coordinates, along with estimated transit properties, are forwarded to 
the follow-up teams who perform the ground-based observation campaign 
(Section~\ref{sec: FU}). A dedicated analysis and transit search is 
performed when the processed data are distributed to the different 
detection teams within the \corot\ Exoplanet Science Team (typically 
several months after the end of the \corot\ observation). 

The detection teams benefit from the total temporal length of the 
lightcurves (about $130$~days for LRa01) and the reduced noise after full 
processing, and are well equipped for finding shallow transits around 
highly variable and active stars. The detection teams try to sort 
out as many false positives as possible in order to avoid the waste 
of observation time and telescope resources on non-planetary objects
(Section~\ref{sec: FU}). See \citet{Cabrera2009} for a full 
description of the \corot\ detection pipeline, with the generation of 
an initial candidate list and subsequent vetting for the presence of 
false alarms arising from eclipsing binaries. In the end, a prioritised 
list of planet candidates and their parameters is composed, which is 
subject to further follow-up observations.

The list of the LRa01 candidates is given in Table~\ref{Tab:all_planets},
along with coordinates, transit period, epoch, duration and depth. Each 
candidate is described in detail in Section~\ref{sec: candidates}.

\section{Ground-based follow-up observations of the \corot\ candidates}
\label{sec: FU}

\begin{table*}[t]
\begin{center}
\caption{Instruments used for the ground-based \corot\ follow-up observations.}
\label{tab: FU_instruments}
\begin{tabular}{p{7.65cm}|l}
\hline
\hline
\noalign{\smallskip}
Instrument & Observatory \\
\noalign{\smallskip}
\hline
\noalign{\smallskip}
\multicolumn{2}{c}{Photometry}\\
\noalign{\smallskip}
\hline
\noalign{\smallskip}
CAMELOT on the IAC 80\,cm telescope; SD camera on the ESA Optical ground-station (ESA-OGS); FastCam at the Carlos Sanchez Telescope (CST)  & Observatorio del Teide, Canary Islands, Spain\\
CCD cameras on the 0.46\,m and 1\,m Wise telescopes & Wise Observatory, Israel\\
EulerCam on the Euler 1.2\,m telescope & La Silla observatory, Chile\\
MegaCam on the 3.6\,m Canada-France-Hawaii telescope (CFHT) & CFH observatory, Hawaii, USA\\
CCD camera on the 1.2\,m MONET-North telescope & McDonald Observatory, Texas, USA\\
NACO on the ESO's VLT & Paranal Observatory, Chile\\
BEST on the DLR 20\,cm telescope & Observatoire de Haute-Provence (OHP), France\\
BEST\,II on the DLR 25\,cm telescope & Observatorio Cerro Armazones, Chile\\
\noalign{\smallskip}
\hline
\noalign{\smallskip}
\multicolumn{2}{c}{Spectroscopy}\\
\noalign{\smallskip}
\hline
\noalign{\smallskip}
AAOmega on the 3.9\,m AAO telescope & Anglo-Australian Observatory, Australia\\
FLAMES-GIRAFFE and UVES on ESO's VLT & Paranal Observatory, Chile\\
CRIRES on ESO's VLT & Paranal Observatory, Chile\\
CORALIE on the 1.2\,m Euler telescope, HARPS on the ESO 3.6\,m  telescope & La Silla observatory, Chile\\
SANDIFORD on the 2.1\,m Otto Struve telescope & McDonald Observatory, Texas, USA\\
Coud\'e spectrograph on the 2\,m TLS telescope&  Th\"uringer Landessternwarte (TLS), Tautenburg, Germany\\
SOPHIE on the 1.93\,m telescope & Observatoire de Haute-Provence (OHP), France\\
HIRES on the 10\,m Keck~I telescope & Keck observatory, Hawaii,USA\\
FIES on the 2.56\,m NOT telescope & Roque de los Muchachos Observatory, Canary Islands, Spain\\
\noalign{\smallskip}
\hline
\end{tabular}
\end{center}
\end{table*}

The detection of planetary transit candidates in the \corot\ 
lightcurves is only the start on a long, and time consuming 
\emph{road} to the successful confirmation of the planetary 
nature. The \corot\ follow-up is done with ground based 
facilities, following the strategy outlined in 
\citet{Alonso2004}. Ground-based follow-up observations are 
motivated by the need to exclude false positives from the 
list of candidates and to obtain a full characterisations of 
the detected planets. Practically all false positives in 
transit searches are due to some configuration involving 
eclipsing binary systems, with the large majority of 
false positives caused by either grazing or diluted 
eclipsing binaries (Section~\ref{sec: Spectroscopy}). Their 
nature was first described by \citet{Brown2003}, with a 
development for CoRoT candidates given by \citet{Almenara2009}.

Photometric observations are needed to confirm that the 
transit is occurring `on target' (Section~\ref{sec: Photometry}). 
If confirmed, complementary RV observations are performed to 
definitely assess the planetary nature of the transiting body and 
eventually reveal the true mass of the planet. High 
S/N ratio spectroscopy at high-resolution are also conducted 
to derive the photospheric fundamental parameters of the planet 
host star and determine its mass and radius 
(Section~\ref{sec: Spectroscopy}).

The \corot\ follow-up program is challenging both in terms of 
telescope capabilities and observation time. The \corot\ follow-up 
community uses time allocated at different observatories as 
described in Sections~\ref{sec: Photometry} and 
\ref{sec: Spectroscopy}. Many RV observations in 2008 and 2009 were 
dedicated to \corot-7 in order to confirm the existence and nature 
of the companions and to constrain their parameters 
\citep{Leger2009,Queloz2009,Hatzes2010,Hatzes2011}. Follow-up 
observations for other candidates were therefore limited or delayed 
to the 2009/2010 and 2010/2011 observing season. 
Table~\ref{tab: FU_instruments} lists all facilities used for the 
follow-up of the LRa01 candidates, Table~\ref{Tab:all_planets_cont} 
lists the results of the follow-up in a concise form. Details can 
be found in the description of the individual candidates in 
Section~\ref{sec: candidates}.

\subsection{Photometry}
\label{sec: Photometry}

The first step in the follow-up sequence is the ground-based 
photometric observation of the transit. It needs to be verified 
that the detected transit signals occur on the main target inside 
the \corot\ photometric mask (typically $20\arcsec$ large). The star 
is observed during a transit, and again between two transits. The 
stellar brightness, as well as that of any other nearby star is 
monitored. Contaminating eclipsing binaries (CEBs) are sorted out 
by this procedure. The method is described in more detail in 
\citet{Deeg2009}.

Required for these observations is the correct and precise
prediction of the epoch of the transit occurrence. Timing errors 
of more than a few hours make the follow-up of transit events 
unfeasible. The ephemeris errors of faint or shallow candidates 
(listed in Table~\ref{Tab:all_planets}) imply that their 
follow-up has to be performed within 1-2 years after \corot\ 
on-board observation.

As part of the photometric follow-up program of \corot, the BEST 
telescope at OHP performed a survey of variable stars in the LRa01 
field prior to the satellite launch \citep{Kabath2008}. The 
eclipsing binaries LRa01~E1~1574 and LRa01~E1~0622 
(Table~\ref{Tab:Binaries}) are among the targets previously found 
in the aforementioned paper.

\subsection{Spectroscopy}
\label{sec: Spectroscopy}

The conclusion on the nature of some transiting objects is 
drawn from complementary time-series RV measurements as well as 
high-resolution, high S/N ratio spectroscopy. 

RV measurements are required to reject possible false-positives 
and confirm the planetary nature of the transiting object. Binary 
system, eventually identified by RV measurements, are classified 
into the following categories: i) binaries with only one stellar 
component spectroscopically visible (SB1); ii) binaries with two 
or more stellar components spectroscopically visible (SB2, SB3, 
etc.); iii) blended eclipsing binary, i.e., spatially unresolved 
eclipsing binary whose light is diluted by the main \corot\ 
target (blend). 

Transits provide the direct measurement of the planet-to-stellar 
radius ratio ($R_P/R_*$), whereas RV measurements yield the mass 
function of the star/planet system. Stellar radii and masses are 
thus needed to determine radii and masses of the transiting 
candidates. A first-oderder estimate of the size of  the transiting 
objects is derived from the spectral types of the host-stars, as 
listed in the Exo-Dat data-base. However, as already described in 
Section~\ref{sec: field}, the photometrically spectral 
classification reported in Exo-Dat suffers some uncertainties. 
High-resolution, high S/N ratio spectroscopy is thus necessary to derive 
stellar mass and radius and, eventually, determine the mass, radius, 
and bulk density of the confirmed planets.

The spectroscopic follow-up observations of LRa01 started with 
a first spectroscopic ``snap-shot'' of some of the \corot\ 
candidates. Low-resolution ($R\approx1\,300$) reconnaissance 
spectroscopy  was performed with the AAOmega multi-fiber 
spectrograph mounted at the 3.9\,m telescope of the Australian 
Astronomical Observatory (AAO) during two observing runs, in 
January 2008 and from December 2008 to January 2009. Further 
multi-object spectroscopic observations were performed with 
the FLAMES-GIRAFFE facility ($R\approx26\,000$) at the ESO 
Very Large Telescope (Paranal Observatory, Chile) in winter 
2005 \citep{Loeillet2008,Gazzano2010}.

The AAOmega and FLAMES-GIRAFFE observations classified the 
stars and derived a fist estimate of their photospheric 
parameters as described in \citet{Gandolfi2008} and 
\citet{Gazzano2010}. If the host star turned out to be a giant 
star, although it was listed as a main sequence star in the 
Exo-Dat database, the size of the transiting object was 
re-evaluated based on the new stellar parameters. Planned 
ground-based photometry and RV follow-up observations were 
cancelled if the size of the transiting body was inconsistent 
with a planetary object. This spectroscopic screening singled 
out also B-type stars and rapidly rotating targets for which 
high-precision RV measurements are not feasible.

The nature of the transiting objects is further investigated through 
reconnaissance high-resolution spectroscopy. This is performed using 
the CORALIE spectrograph at the 1.2\,m Euler telescope in La Silla 
observatory, the Sandiford cassegrain echelle spectrograph on the 
2.1\,m telescope at McDonald Observatory, and the coud\'e echelle 
spectrograph of the 2\,m telescope of the Th\"uringer Landessternwarte 
(TLS). If this stage is successfully passed, the planetary candidate 
is handed down to SOPHIE at the 1.93\,m telescope at the Observatoire 
de Haute-Provence (OHP). In its high efficiency mode ($R\approx 40\,000$), 
SOPHIE is able to reach RV precision of a few dozen \ms\ on a 
solar-like star, down to $V\approx 14.5$\,mag. This accuracy is 
fully sufficient for detecting Jupiter-like, and even Saturn-like 
planets, in a close-in orbit around a solar-like star. We recently 
took also advantage of the fiber-fed FIES spectrograph attached at 
the 2.56\,m Nordic Optical Telescope (NOT). The recent 
refurbishments carried out at this instrument had improved the 
capability of FIES for very high-precision RV measurements down 
to $\sim10$\,\ms, making this spectrograph a precious resource to 
use for \corot\ RV follow-up.

The final ``step'' of the RV follow-up uses high-precision RV 
measurements with HARPS at ESO's 3.6\,m telescope. The RV follow-up 
of the faintest candidates in LRa01 ($15\lesssim V \lesssim16$\,mag) 
has been strengthened using the HIRES echelle spectrograph on the 
10\,m Keck I telescope.

The photospheric parameters of the candidates, i.e., effective 
temperature (\teff), gravity (\logg), metallicity ($[M/H]$), and  
projected rotational velocity (\vsini), are usually derived by 
analysing also the acquired high-resolution spectra, as already 
described in other \corot\ exoplanet papers 
\citep[e.g., ][]{Deleuil2008,Leger2009,Bruntt2010,Gandolfi2010}. 
Stellar masses and radii are then inferred by comparing the location 
of the objects on a \teff\,$vs.$\,\logg\ diagram with theoretical 
evolutionary tracks.
Only an estimate of the spectral type can be derived for 
candidates with low S/N ratio spectra ($<$10-15), as described in some 
cases in Section~\ref{sec: candidates}.  

For the confirmed planetary candidates high S/N ratio spectra are 
usually acquired with HARPS, HIRES, and UVES (ESO, Paranal 
Observatory) spectrographs.

\section{\corot\ planetary candidates}
\label{sec: candidates}

Here we present the \corot\ transit candidates along with a brief 
overview over the properties of the star and the detected transit 
signal. Any follow-up observations that have been performed 
are also described. The candidates are presented in the following 
order: confirmed planets, identified non-planetary objects, unsettled 
good planetary candidates, unsettled low priority planetary candidates 
(suspected binaries), false alarms, and the so-called ``X-case" 
candidates (see below).

A planet is considered confirmed when RV-measurements definitely assess 
the planetary nature of the transiting object and allow to determine 
its mass (Section~\ref{Sec:confirmed planets}). Candidates listed as 
``settled cases - non planetary objects" (Section~\ref{Sec:Settled cases}) 
are objects whose non planetary nature were identified by either 
photometric or spectroscopic follow-up observations. These include 
blends, contaminating eclipsing binaries (CEB), and binary systems.

Most of the observing time reserved for LRa01 was invested on \corot-7. 
Therefore many candidates were being followed-up in the 2009/2010 
and some even in the 2010/2011 observing seasons. Still, follow-up 
observations could not be concluded for all candidates. These are 
listed as unresolved cases and are subdivided into ``unsettled good 
planetary candidates" (Section~\ref{Sec: Unsettled good candidates}) 
and ``unsettled low priority planetary candidates" 
(Section~\ref{Sec: Unsettled low priority candidates}), based on the 
analysis of the \corot\ lightcurves only. The latter were usually not 
followed-up due to one or more bad characteristics hinting at
stellar binary scenario (e.g., out of transit variations, depth 
differences between even and odd transits or different color channels, 
very shallow secondary eclipse). 

For the sake of completeness ``false alarm" objects are also included 
(Section~\ref{Sec: False alarms}). These are shallow transit candidates 
that were identified by one detection team in lightcurves heavily 
affected by instrumental effects. They are considered as probable false 
alarms because they could not be reproduced by other detection teams 
using different filtering techniques. The ``X-case'' candidates 
(Section~\ref{Sec: X-cases}) are objects that might be planetary 
candidates if the spectral type of the target star were considerable 
different than the one listed in Exo-Dat. They have a very low priority 
in the follow-up program and were not observed so far. Finally, in some 
cases only a single transit event is present in the \corot\ lightcurve. 
Those are listed as ``mono-transits'' (Section~\ref{Sec: Mono-transits}). 
The depth of the signals indicates that these are eclipsing stellar 
binaries (Table~\ref{Tab:Binaries}).

A concise list of the transit parameters and the follow-up status of 
the candidates can be found in Tables~\ref{Tab:all_planets} and 
\ref{Tab:all_planets_cont}, respectively. The RV-measurements performed 
on the \corot\ LRa01 candidates are listed in Table~\ref{rvtable}. For 
the RV data of the confirmed exoplanets in LRa01, we refer the reader 
to the respective articles reporting on their discovery. 

\subsection{Confirmed planets}
\label{Sec:confirmed planets}

\begin{table*}
\begin{center}
\caption{Parameters of the planets detected in the \corot\ LRa01 field.}
\label{tab: planetary_param}
\begin{tabular}{cccccp{3cm}}
\hline
\hline
\noalign{\smallskip}
Planet & Host star spectral type & Planetary mass & Planetary radius & Semi-major axis [AU] & Source \\
\noalign{\smallskip}
\hline
\noalign{\smallskip}
~\,\corot-7b     & G9V & $4.8\pm0.8\,M_{Earth}$              & $1.68\pm0.09\,R_{Earth}$           & $0.0172\pm0.0002$               & \citet{Leger2009,Queloz2009} \\
                 & G9V & $6.9\pm1.5\,M_{Earth}$              & $1.68\pm0.09\,R_{Earth}$           & $0.0172\pm0.0002$               & \citet{Leger2009,Hatzes2010} \\
                 & G9V & $5.2\pm0.8\,M_{Earth}$              & $1.58\pm0.10\,R_{Earth}$           & $0.0172\pm0.0002$               & \citet{Queloz2009,Bruntt2010}\\
                 & G9V & $2.3\pm1.8\,M_{Earth}$              & $1.58\pm0.10\,R_{Earth}$           & $0.0172\pm0.0002$               & \citet{Pont2011,Bruntt2010}  \\
                 & G9V & $8.0\pm1.2\,M_{Earth}$              & $1.58\pm0.10\,R_{Earth}$           & $0.0172\pm0.0002$               & \citet{Ferraz2011,Bruntt2010}\\
                 & G9V & $5.7\pm2.5\,M_{Earth}$              & $1.58\pm0.10\,R_{Earth}$           & $0.0172\pm0.0002$               & \citet{Boisse2011,Bruntt2010}\\
                 & G9V & $7.42\pm1.21\,M_{Earth}$            & $1.58\pm0.10\,R_{Earth}$           & $0.0172\pm0.0002$               & \citet{Hatzes2011,Bruntt2010}\\
\noalign{\smallskip}
\hline
\noalign{\smallskip}
~\,\corot-7c$^*$ & G9V & $~\,8.4\pm0.9\,M_{Earth}\sin i$     & -                                  &               $0.046$           & \citet{Queloz2009}\\
                 & G9V & $12.4\pm0.4\,M_{Earth}\sin i$       & -                                  &               $0.045$           & \citet{Hatzes2010}\\
                 & G9V & $13.6\pm1.4\,M_{Earth}\sin i$       & -                                  &               $0.045$           & \citet{Ferraz2011}\\
                 & G9V & $13.2\pm4.1\,M_{Earth}\sin i$       & -                                  &               $0.045$           & \citet{Boisse2011}\\
\noalign{\smallskip}
\hline
\noalign{\smallskip}
~\,\corot-7d$^*$ & G9V & $16.70\pm0.42\,M_{Earth}\sin i$     & -                                  &               $0.080$           & \citet{Hatzes2010}\\
\noalign{\smallskip}
\hline
\noalign{\smallskip}
~\,\corot-5b     & F9V & $0.467^{+0.047}_{-0.024}\,M_{Jup}$  & $1.388^{+0.046}_{-0.047}\,R_{Jup}$ & $0.04947^{+0.00026}_{-0.00029}$ & \citet{Rauer2009}\\
\noalign{\smallskip}
\hline
\noalign{\smallskip}
\corot-12b       & G7V & $0.917\pm0.07\,M_{Jup}$             & $~\,1.44\pm0.13\,R_{Jup}$          &        $0.0402\pm0.0009$        & \citet{Gillon2010}\\
\noalign{\smallskip}
\hline
\end{tabular}
\end{center}
\vspace{-0.2cm}
$^*$ \corot-7c and \corot-7d were detected by RV measurements only. The real nature of \corot-7d of is not 
definitely assessed. See \citet{Patzold2011} for the parameters of CoRoT-21b.
\end{table*}

Four candidates detected during the \corot\ LRa01 run were confirmed 
as \emph{bona fide} transiting planets (see Table~\ref{tab: planetary_param} 
for the full parameters). Three Jupiter-sized planets, \corot-5b 
\citep{Rauer2009}, \corot-12b \citep{Gillon2010}, and \corot-21b 
\citep{Patzold2011}, and the first transiting terrestrial planet 
\corot-7b \citep{Leger2009, Queloz2009}. For the very first time, 
the bulk density of a small extrasolar planet was derived 
consistent with the bulk densities of terrestrial planets  (see 
Table~\ref{tab: planetary_param}). An accurate reanalysis of the 
acquired HARPS and UVES spectra was recently published by 
\citet{Bruntt2010}, leading to an improvement of stellar parameters 
of \corot-7.

There are now a number of independent determinations of the mass of 
CoRoT-7b. See Table~\ref{tab: planetary_param} and \citet{Hatzes2011} 
for a detailed discussion. We note that CoRoT-7b is almost identical to 
the recently discovered transiting Super-Earth Kepler-10b. Kepler-10b 
has a period of about $P=0.84$ days, mass $M_P=4.56^{+1.17}_{-1.29}~M_{Earth}$, 
radius $R_P=1.416^{+0.033}_{-0.036}~R_{Earth}$, and mean density 
$\rho=8.8^{+2.1}_{-2.9}$\,\gcm3 \citep{Batalha2011}.

\corot-7b is the only transiting planet in the \corot-7b system. Two other 
planets are inferred from RV measurements only: \citet{Queloz2009} 
discovered \corot-7c; \citet{Hatzes2010} found evidences of the presence 
of a third planet, \corot-7d (see also Table~\ref{tab: planetary_param}), but 
further RV measurements are required to definitely assess its planetary nature.
The detection of these two additional planets is disputed by \citet{Pont2011}. 

\begin{figure}[th]
  \centering
  \begin{minipage}[b]{0.84\linewidth}
    \includegraphics[%
      width=0.9\linewidth,%
      keepaspectratio]{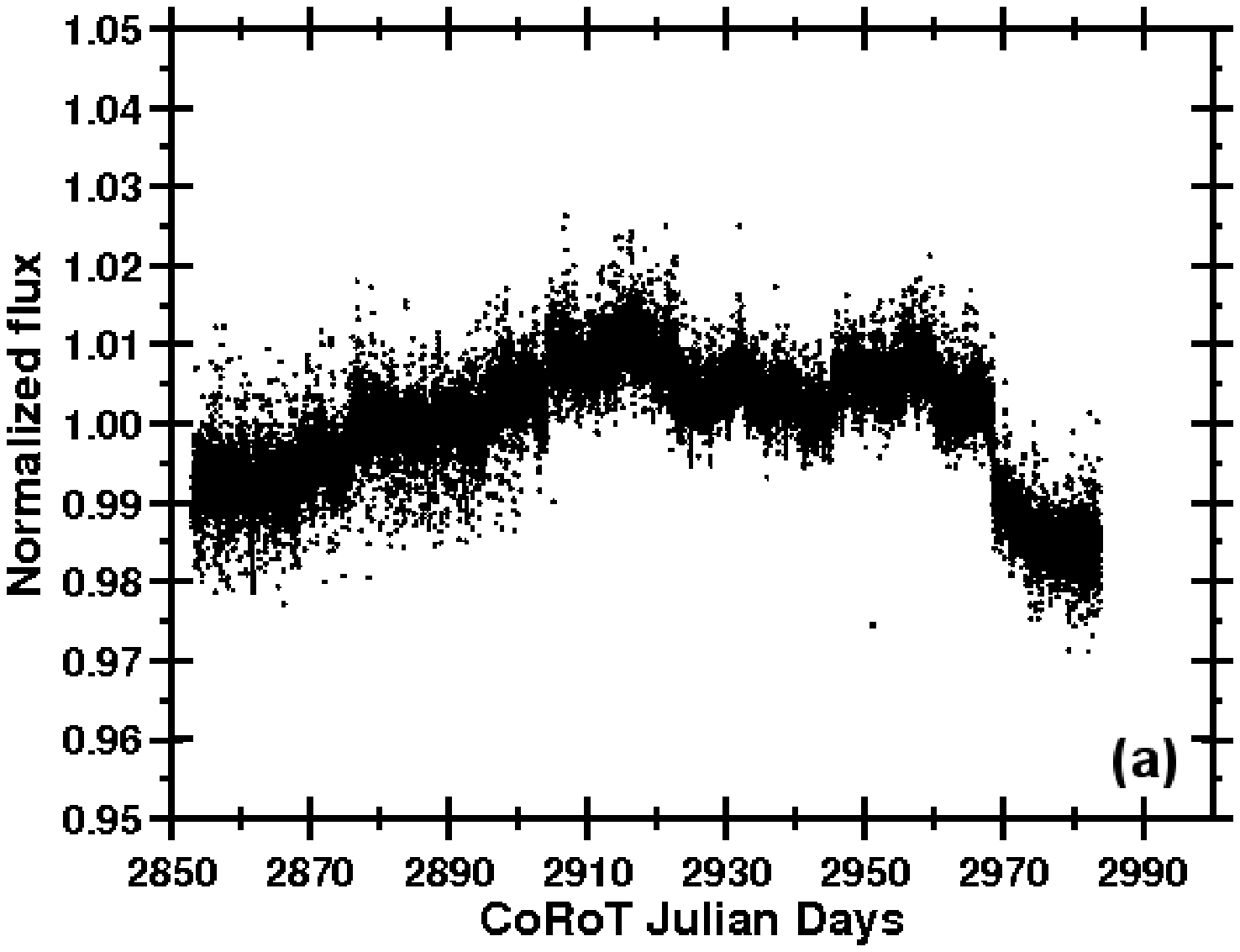}
      \end{minipage}

  \begin{minipage}[b]{0.84\linewidth}
    \includegraphics[%
      width=0.9\linewidth,%
       keepaspectratio]{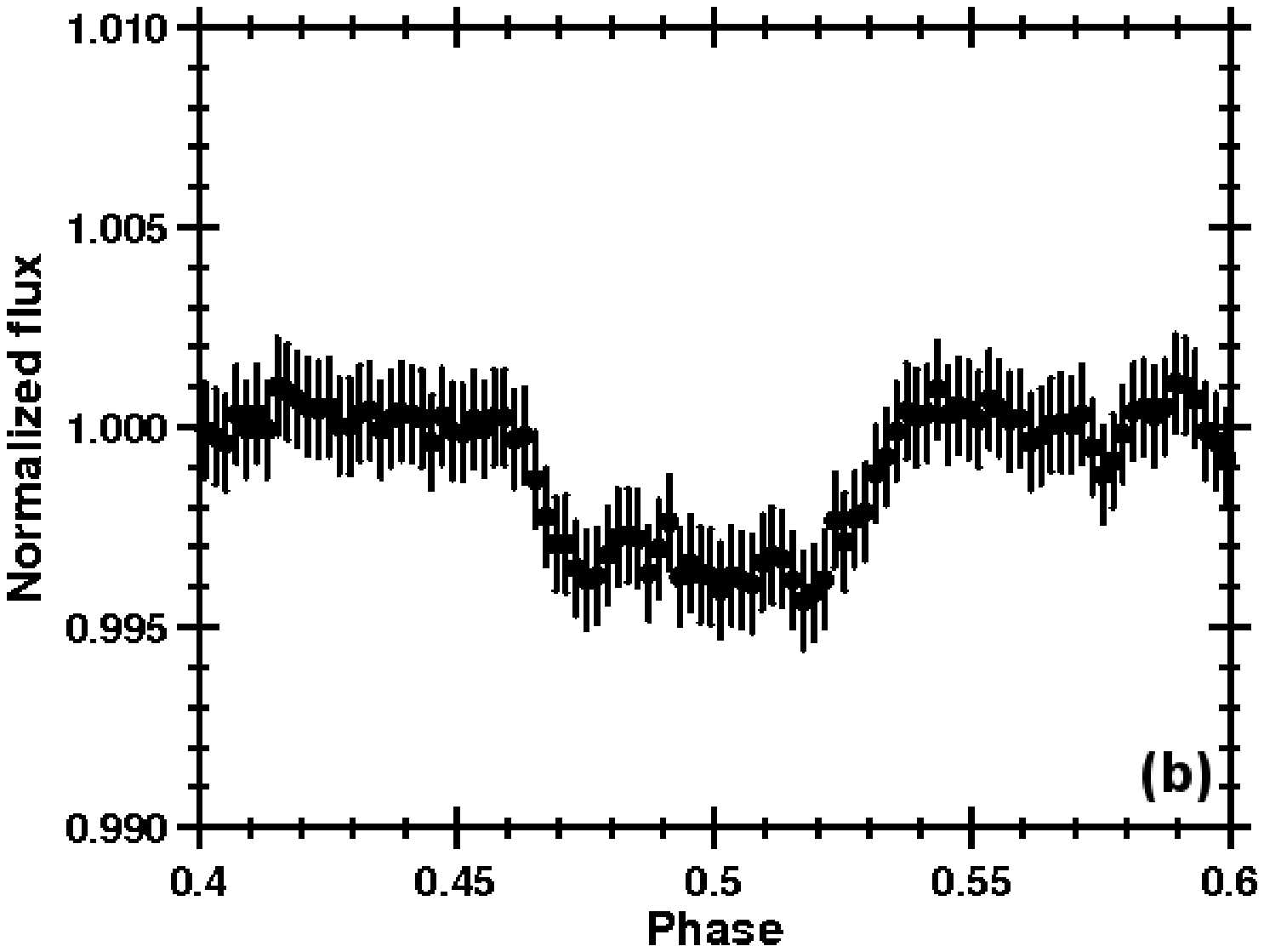}
      \end{minipage}
  \caption{
  \emph{Top panel}: Raw white lightcurve of LRa01~E2~5277 (also known as \corot-21b). 
  \emph{Bottom panel}: Normalized and phase-folded raw white lightcurve of the transit 
  candidate LRa01~E2~5277 at the transit period $P=2.73$ days. A baseline fit has been 
  subtracted around each transit before folding.}
  \label{fig:E2_5277}
\end{figure}

CoRoT-21b, also known as candidate LRa01\,E2\,5277 (\corot\ ID 0102725122), 
has been recently confirmed as a transiting hot-Jupiter planet. The lightcurve 
of its faint host star ($V=16.1$\,mag) contains a 0.45\,\% deep transit signal 
with a period of 2.73~days (Figure~\ref{fig:E2_5277}). Photometric follow-up 
observations performed with IAC80 confirm the transit event on target. RV 
measurements conducted with HARPS, and recently with HIRES, show significant 
variations in phase with the \corot\ ephemeris and consistent 
with a $\sim$$2~M_{Jup}$ planet around a F8\,IV star (\teff$\approx$6100\,K, 
\logg$\approx$3.5\,dex). Since the parameters of CoRoT-21b are still under 
investigation, the planet is not listed in Table~\ref{tab: planetary_param} 
but in Tables~\ref{Tab:all_planets} and \ref{Tab:all_planets_cont}. It will 
be presented in a forthcoming paper \citep{Patzold2011}, along with the HARPS
and HIRES RV measurements.

\subsection{Settled cases: non planetary objects}
\label{Sec:Settled cases}

The following objects are identified as non-planetary objects based on 
ground-based follow-up observations.

\subsubsection{LRa01 E1 0544 - CHR - 0102714746}

LRa01~E1~0544 is a relatively bright star ($V=13.39$\,mag) with a 
0.15\,\% deep eclipse occurring every 2.75 days. AAOmega 
observations classify the target as a F7 dwarf star, in good 
agreement with the  classification listed in Exo-Dat (F8\,IV).
SOPHIE spectroscopic observations indicate a fast rotator with 
no significant RV variations down to a precision of 50~m\,s$^{-1}$. 
EulerCam measurements show no transit events on target. Instead a 
$\sim$4 magnitude fainter star located 9\arcsec~West of the main 
target has deep eclipses ($D$$\approx$20\,\%). IAC80 observations 
confirm this result. This object is a contaminating eclipsing 
binary (CEB).

\subsubsection{LRa01 E1 0561 - CHR - 0102597681}

This is a relatively bright candidate ($V=12.00$\,mag) listed
as a SpT=A0\,V star in Exo-Dat. The \corot\ lightcurve shows a 
0.70\,\% deep transit occurring every 20.82 days superimposed 
on a $\gamma$-Doradus like pulsations. Low-resolution AAOmega 
spectroscopy classifies this object as a A7\,IV/V star. Two 
moderate S/N$\approx35$ ratio SOPHIE spectra reveal a low-contrast 
single peak cross-correlation function (CCF) with a RV variation 
of about 52~km\,s$^{-1}$, in anti-phase with the \corot\ 
ephemeris, i.e., the eclipses occur on the rising part of the 
RV curve and are caused by the star whose CCF peak is detected
in the SOPHIE spectra. A single epoch UVES spectrum with higher 
S/N ratio ($\sim120$) unveils the presence of other two components in 
the system, making the candidates a spectroscopically resolved 
SB3 system with pulsating components.

\subsubsection{LRa01 E1 2890 - MON - 0102618931}

According to Exo-Dat, this candidate has an apparent 
V-magnitude of 15.73 and its spectral type is G5\,III.
IAC80 on-off photometry shows that a contaminating eclipsing 
binary is the origin of the transit signal in this lightcurve 
($D=0.29$\,\%, $P=2.43$~days). It is $\sim3.2$\,mag fainter than 
the target and positioned $\sim12$\arcsec\ South-East from 
LRa01~E1~2890.

\subsubsection{LRa01 E1 3666 - MON - 0102790970}

This is a faint ($V = 15.47$\,mag) F5\,V star (Exo-Dat) with an 
apparent transit signal of 0.45\,\% depth and a period of 1.55 days.
CFHT and IAC80 observations identify a nearby background star 
$\sim8$\arcsec\ West of the target with a variation of 1.5\,\% as 
the source of the signal. The candidate is a contaminating eclipsing 
binary (CEB).

\subsubsection{LRa01 E1 5015 - MON - 0102777869}

This is a candidate around a relatively faint ($V=16.17$\,mag) 
G2\,V star (Exo-Dat) with a 1\,\% deep transit signal and a period 
of 13.69 days. The long transit duration of about 10 hours implies 
an eclipsing binary.

\begin{figure}
\includegraphics[%
      width=0.8\linewidth,%
      keepaspectratio]{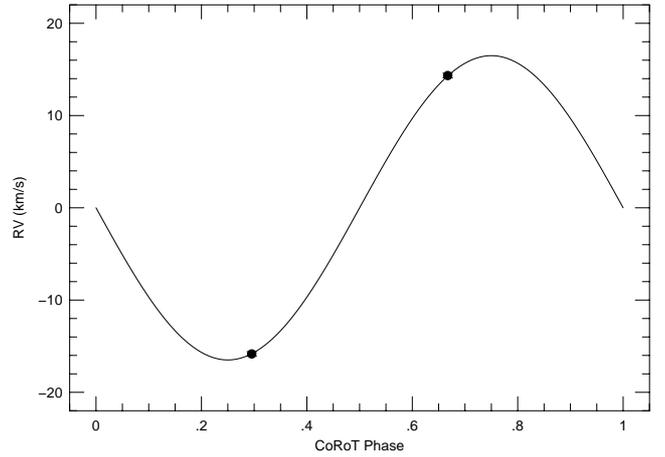}
        \caption{The HARPS RV measurements of LRa01\,E1\,5015 (points) 
	         phased to the \corot\ transit period ($P=13.69$~days) 
		 and epoch. The best fitting sine curve is over-plotted 
		 with a line. Note that the systemic velocity has been 
                 subtracted.}
  \label{fig:RV_E1_5015}
\end{figure}

ESA-OGS observations show that the transit is on target. 
Figure~\ref{fig:RV_E1_5015} reports the HARPS RV measurements of 
LRa01~E1~5015 phase folded to the \corot\ transit period and epoch, 
along with a sine fit. The semi-amplitude is $K=16.5$~km\,s$^{-1}$ which 
corresponds to a mass function\footnote{We remind the reader 
that the mass function expressed in solar units is defined as 
$f(m)=M_{2}^{3}\,(sin^{3}i)/(M_{1}+M_{2})^{2}=
(1.03608\times10^{-7})(1-e^{2})^{3/2}\,K_{1}^3\,P$, where $M_{1}$ 
and $M_{2}$ are the masses in solar units of the primary and secondary 
component, $i$ and $e$ the orbit inclination and eccentricity, $K_{1}$ 
the RV semi-amplitude of the primary star in km\,s$^{-1}$, and $P$~the 
orbital period expressed in days.} $f(m)=0.00637$~$M_\odot$. Assuming 
a mass of 1\,$M_\odot$ for the main component, the companion mass is 
$\approx0.18\,M_\odot$. This is a SB1 system with a low-mass companion 
star.

\subsubsection{LRa01 E1 4353 - MON - 0102692038}

This candidate in the lightcurve of a $V=15.78$\,mag star of spectral 
type A5\,V (Exo-Dat) has 1.09\,\% deep eclipse with a period of 5.23 
days. EulerCam on-off photometry shows a constant flux in the target 
but a 5\,\% deep eclipse in a contaminant star (\corot\ ID: 0102691690, 
$V=16.7$\,mag). IAC80 confirms the EulerCam observations. This is a 
contaminating eclipsing binary (CEB).

\subsubsection{LRa01 E2 1123 - MON - 0102615551}

A 1.80\,\% deep signal with a period of 3.88~days is detected in 
the \corot\ lightcurve of this candidate. IAC80 and Wise 
photometry confirms the transit signal is on target. UVES and HARPS 
observations of this $V=14.62$\,mag object identify the target as a 
K5 dwarf star instead. Two CORALIE and six HARPS RV measurements show 
no significant sinusoidal variations with an amplitude greater than 
$\sim$50\,\ms.

An examination of the Ca\,{\sc ii}~H \& K region from spectra taken 
with UVES shows three emission components. The RVs of two components 
vary in phase with twice the transit period and with a maximum 
velocity difference of about 67~km\,s$^{-1}$. This is a blend scenario:
a probable hierarchical triple system consisting of a K5 primary 
active dwarf orbited by two eclipsing active M-type stars that are too 
faint to be seen in the metallic lines used for the RV measurements.

\subsubsection{LRa01 E2 1145 - CHR - 0102707895}

A transit with the following parameters is detected in the \corot\ 
lightcurve: $D=0.43$\,\% and $P=5.78$~days. The $V=13.96$\,mag 
target is already known from the \corot\ IRa01 run 
\citep[IRa01~E1~1873;][]{Carpano2009} and is classified 
as a A9\,IV/V star on the basis of AAOmega observations. 
SOPHIE measurements spectroscopically resolve the target as 
a SB1 system with a RV curve in anti-phase with the \corot\ 
ephemeris. Assuming a circular orbit, the RV curve 
semi-amplitude of the eclipsing star is K=23.5\,km\,s$^{-1}$.

\subsubsection{LRa01 E2 1897 - MON - 0102658070}

This $V=14.72$\,mag candidate, classified as a F2\,II star 
according to Exo-Dat, shows a deep transit-like signal ($D=2.80$\,\%) 
with a period of 4.67 days. Hints of a secondary eclipse are 
detected in the \corot\ lightcurve at phase=0.5. CFHT photometric 
observations show that the signal originates from an eclipsing 
binary located $\sim3$\arcsec\ Northeast from the \corot\ main 
target. The contaminant exhibits flux variation of about 7.8\,\% 
occurring at the predicted \corot\ ephemeris. This is consistent 
with the transit signal when the dilutions by the main target and 
by a second brighter star located 24.4\arcsec\ Southwest, are 
considered. This candidate is a contaminating eclipsing binary 
(CEB).

\subsubsection{LRa01 E2 2249 - CHR - 0102755837}

\begin{figure}
    \includegraphics[%
      width=0.9\linewidth,%
      keepaspectratio]{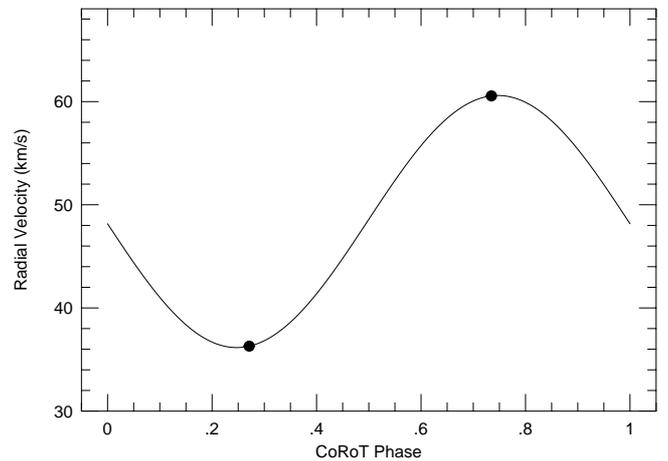}
    \caption{The SOPHIE RV measurements of LRa01\,E2\,2249 (points) 
             with a sine-fit (line) using the \corot\ transit period 
             and epoch.}
  \label{fig:RV_E2_2249}
\end{figure}

This candidate has a 0.38\,\% deep transit occurring every 27.93 
days. Low-resolution spectra obtained with AA0mega classify the star 
($V=13.88$\,mag) as K0\,III/IV, in good agreement with the Exo-Dat 
spectral type (K0\,III). SOPHIE RV measurements show the candidate 
to be a SB1 system (K=12~km\,s$^{-1}$). Figure~\ref{fig:RV_E2_2249} 
shows the RV measurements with a sine fit using the \corot\ transit 
period and epoch. The mass function for the system is 
$f(m)=0.0052~M_\odot$. Assuming a primary mass of 1\,$M_\odot$ this 
results in a secondary mass of 0.17~$M_\odot$.

\subsubsection{LRa01 E2 2481 - CHR - 0102723949}

The \corot\ lightcurve of this candidate ($V=13.96$\,mag, 
SpT=F5\,III; Exo-Dat) contains transits with a depth of 1.20\,\% 
and a period of 51.76~days. It is also found in the IRa01 run 
(IRa01\,E1\,2046) and listed as a mono-transit candidate 
in the IRa01 report papers \citep{Carpano2009,Moutou2009}. 
The spectral classification based on low-resolution spectroscopy 
performed with AAOmega hints at a stellar binary. According to 
the method described in \citet{Gandolfi2008}, the spectral type 
of this object ``changes'' from G0\,V to G8\,V as a function 
of the fitted spectral region, suggesting the presence of two 
stellar objects whose lines are blended in the low-resolution 
AAOmega spectrum. A single epoch SOPHIE spectrum shows a SB2 
system \citep{Moutou2009}, confirming the AAOmega binary 
scenario.

\subsubsection{LRa01 E2 2694 - CHR - 0102590741}

Transit events with 1.30\,\% depth and a 30.40 days 
period are found in the lightcurve of this star ($V=13.56$\,mag,
SpT=A0\,IV; Exo-Dat). SOPHIE and HARPS spectra show no CCF 
and only He\,{\sc i} absorption and strong emission Balmer 
lines, indicative of a Be-type star. Low-resolution AAOmega 
spectra confirm the SOPHIE results and yield a spectral type 
of B3Ve which translates into a stellar radius of about 
5~$R_\odot$ \citep{Cox2000}. Taking into account the 
contamination level for this star ($\sim14$\% according to 
Exo-Dat), the observed transit, if on target, is therefore 
caused by a stellar object with a radius of about 
0.6~$R_\odot$.

\subsubsection{LRa01 E2 4129 - MON - 0102590008}
\label{sec:BEB_example}

The $V=15.71$\,mag G0\,IV target star (Exo-Dat) has two 
faint nearby contaminants located about 4.5\arcsec\ to the 
North-Northeast. Wise observations are inconclusive. 
EulerCam photometric observations show no signs of transits 
on the main target. Instead a 7\,\% drop in brightness is 
found in one of the two contaminant stars which accounts for 
the transit events observed on LRa01\,E2\,4129 (0.18\,\% deep 
eclipses every 1.94 days). Therefore, this case is classified 
as a contaminating eclipsing binary (CEB).

\subsubsection{LRa01 E2 5084 - MON - 0102667981}

This $V=15.95$\,mag transit candidate is classified 
as a A5 sub-giant star according to Exo-Dat. The transit is 
0.28\,\% deep with a 9.92 days period. Based on HARPS 
observations, the candidate is a SB1 system. Assuming a 
circular orbit, the HARPS measurements yield a 
$K=37.2$~km\,s$^{-1}$ RV curve in anti-phase with the 
\corot\ ephemeris.

\subsubsection{LRa01 E2 5184 - CHR - 0102779966}

\begin{figure}[t]
    \includegraphics[%
      width=1.0\linewidth,%
      keepaspectratio]{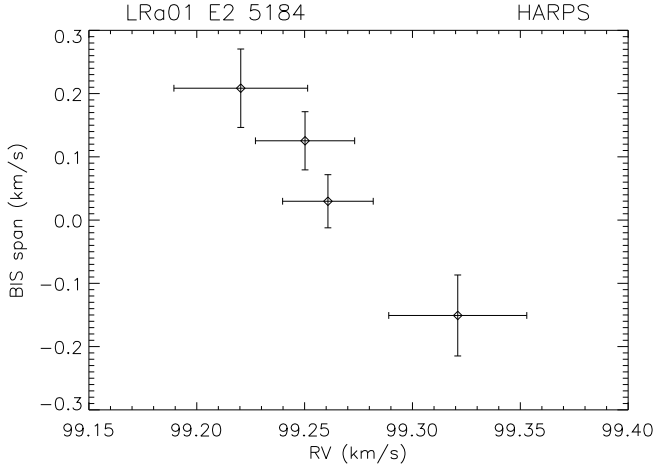}
  \caption{CCF bisector spans versus RV measurements as derived 
           from the HARPS spectra of LRa01\,E2\,5184.}
  \label{fig:Bis_E2_5184}
\end{figure}

This V=15.41\,mag candidate ($D=0.41$\,\%, $P=7.37$~days),
classified as K2\,V star in Exo-Dat, is already known from 
the IRa01 field as IRa01\,E1\,4108 \citet{Carpano2009,Moutou2009}. 
CFHT ground-based photometry confirms the transit to be on target. 
HARPS spectra yield \teff\,$=5000\pm100~K$, \logg\,$=4.4\pm0.1$\,dex,
$[M/H]=0.07\pm0.06$\,dex, \vsini\,$=1.5\pm0.5$\,km\,s$^{-1}$, and 
SpT = K0\,V. HARPS data also indicate a strong bisector-RV correlation 
consistent with a blended eclipsing binary (i.e., diluted triple system 
or background/foreground eclipsing binary), as shown in 
Figure~\ref{fig:Bis_E2_5184}.

\subsubsection{LRa01 E2 5747 - MON - 0102753331}
\label{LRa01_E2_5747}

This candidate ($V=16.16$ mag) is already known from the 
IRa01 field as IRa01~E1~4617 \citep{Carpano2009,Moutou2009}. 
The transit is 3.64\,\% deep, appears every 19.75 days, and 
has a duration of 14.13 hours suggesting a stellar companion. 
No CCF is detected with HARPS. The spectrum shows only broad 
Balmer and Mg\,{\sc i}-b lines, indicative of a rapidly 
rotating A-type star, in agreement with the Exo-Dat spectral 
classification (A5\,IV). Even assuming a late A-type dwarf star, 
the stellar radius would be too large \citep[$R_*>1.5~R_\odot$; ]
[]{Cox2000} to make the observed 3.64\,\% deep transit of 
planetary origin. Therefore the transiting object, if on target, 
is a stellar companion.

\subsubsection{LRa01 E2 3739 - MON - 0102755764}

The \corot\ lightcurve of this candidate ($V=15.55$\,mag) 
shows a V-shaped, 2.93\,\% deep transit signal with a duration 
of 6.97 hours and a period of 61.48 days, superimposed on a 
low-amplitude pulsation \citep{Debosscher2009}. Shape, duration, 
and depth suggest grazing eclipses of an evolved star by a 
stellar companion. A single transit event was already observed 
in the IRa01 field (IRa01\,E1\,4014), whereas two transits are 
found in the LRa01 lightcurve. EulerCam observations confirm the 
transit to be on target but 0.14 days after the predicted 
ephemeris. Taking into account the transit timing error of about 
0.07 days at the time of the EulerCam follow-up (25 January 2009), 
the observed transit occurred only $2\sigma$ after the predicted 
event. Nearby stars were either stable during the observation or 
their variation was too small to account for the transit signal 
detected in the \corot\ lightcurve. Therefore the transit is 
concluded to be on target. HARPS observations show no CCF for 
this target. Only broad Balmer lines are visible in the HARPS 
spectrum. This is consistent with the star being a rapidly 
rotating A-type star, in agreement with the Exo-Dat A5\,IV 
spectral type. As described in Section~\ref{LRa01_E2_5747}, 
an eclipsing stellar companion star is suspected to orbit 
around this candidate and follow-up observations are 
completed.

\subsubsection{LRa01 E2 5756 - MON - 0102582529}

A 2.72\,\% deep transit with a period of 15.84 days 
is found in the \corot\ lightcurve of this $V=16.24$\,mag 
candidate, whose spectral type is F0\,V according to Exo-Dat.
The transit signal is rather deep and the ingress/egress 
steeper than expected for a planetary candidate. But it is 
not ruled out as a binary on the photometric level, because 
the transit can also be produced by a planet with a higher 
impact parameter.

IAC80 photometry establises that the transit event is on-target. 
Spectral observations with HARPS reveals no CCF. The spectrum is 
consistent with LRa01\,E2\,5756 being a rapidly rotating A-type 
star. No further follow-up observations are foreseen since the 
transiting object is suspected to be a stellar companion.

\subsection{Unsettled good planetary candidates}
\label{Sec: Unsettled good candidates}

\subsubsection{LRa01 E1 0286 - CHR - 0102742060}

\begin{figure}[t]
  \centering
  \begin{minipage}[b]{0.75\linewidth}
    \includegraphics[%
      width=0.9\linewidth,%
       keepaspectratio]{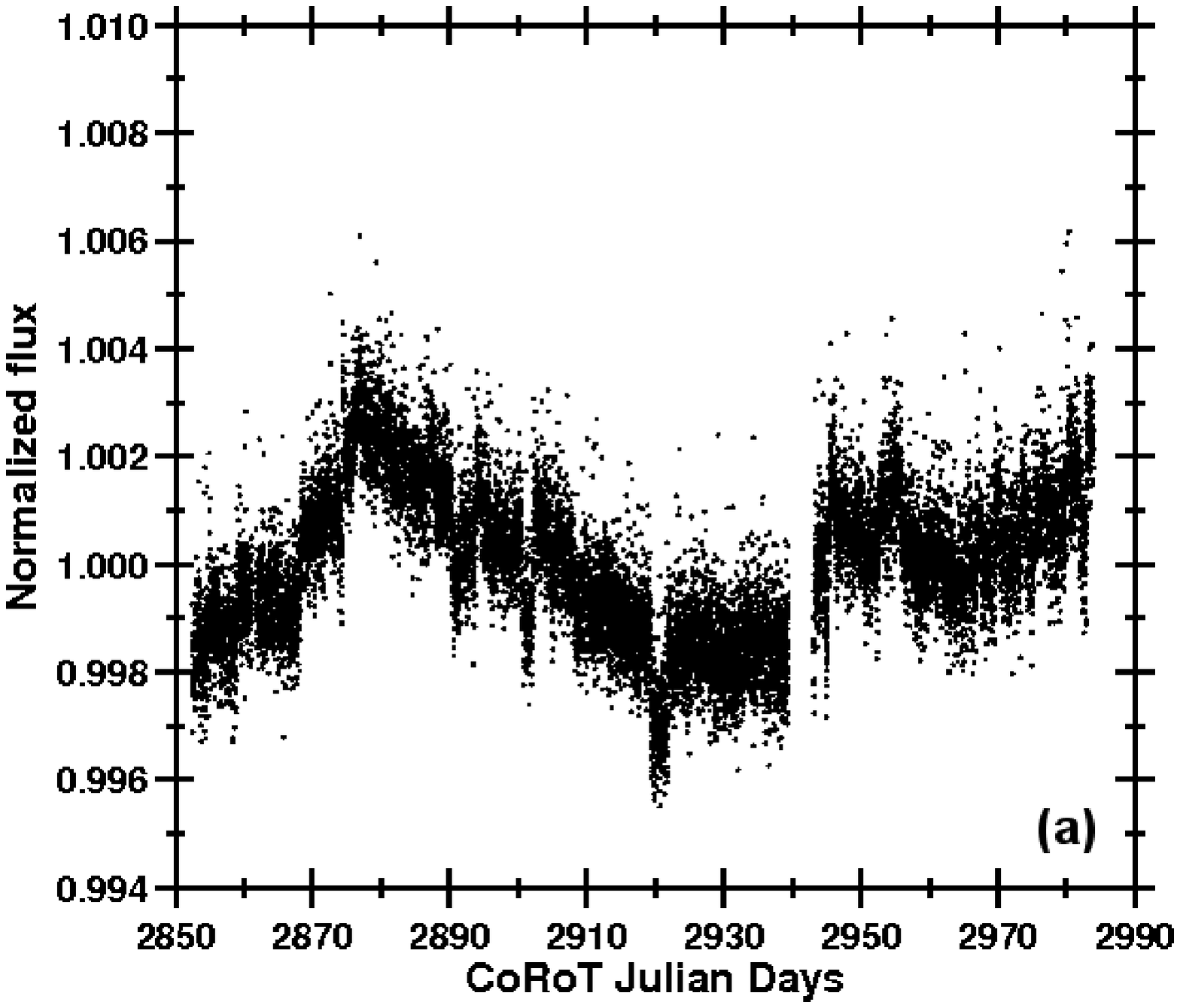}
      \end{minipage}

  \begin{minipage}[b]{0.75\linewidth}
    \includegraphics[%
      width=0.9\linewidth,%
      keepaspectratio]{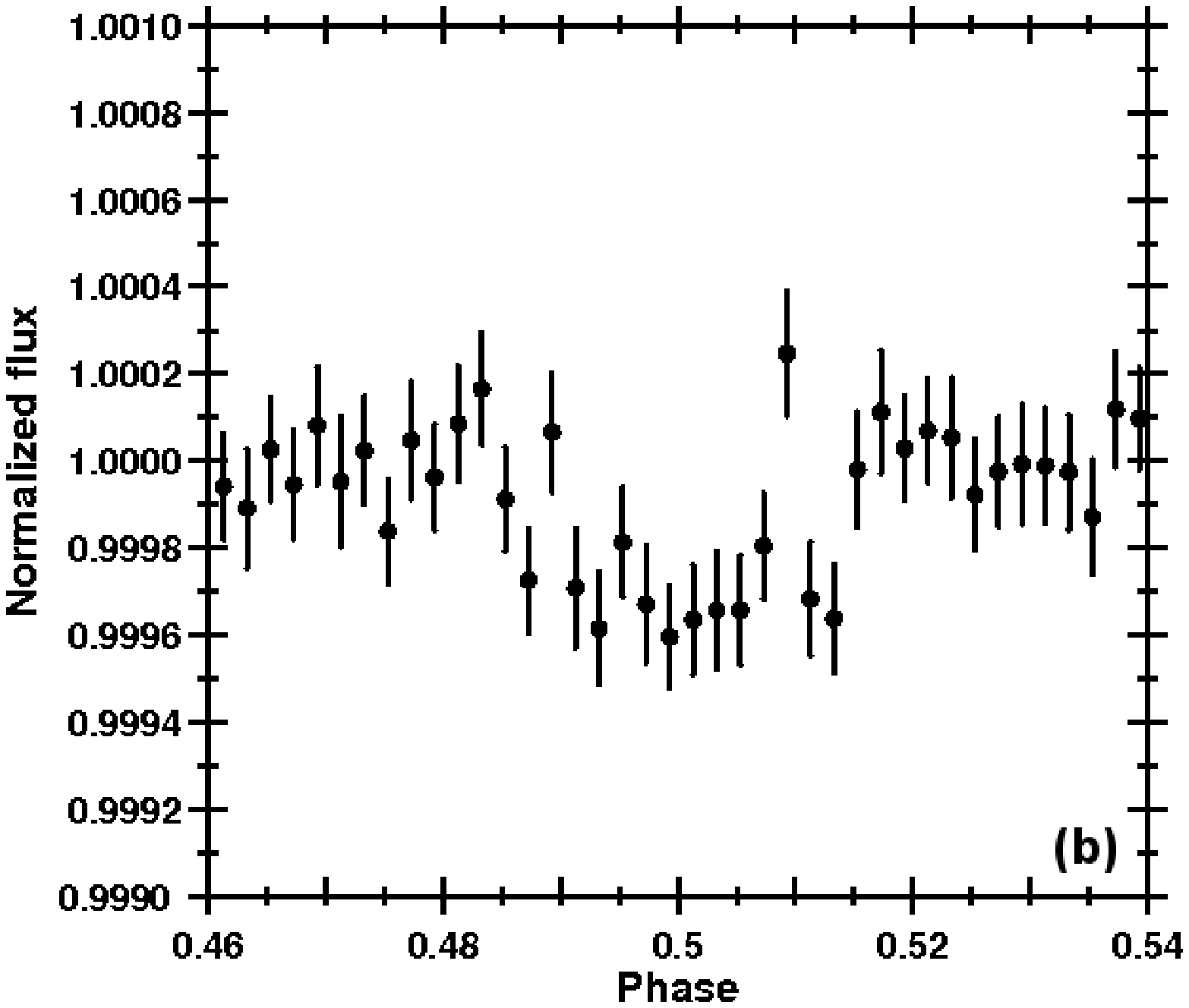}
      \end{minipage}
  \caption{\emph{Top panel}: Raw white lightcurve of LRa01\,E1\,0286 
           which shows instrumental effects (jumps). \emph{Bottom panel}: 
           Normalized and phase-folded white lightcurve of the transit 
           candidate LRa01\,E1\,0286 at the period $P=3.60$ days after 
           filtering with ExoTrans \citep{Grziwa2011}. Additional 
           filtering was necessary to make the transit visible in 
           this example.}
  \label{fig:E1_286}
\end{figure}

A shallow eclipse (0.03\,\%) in the lightcurve of this relatively bright 
star $V=13.30$\,mag was discovered at a period of 3.60~days 
(Figure~\ref{fig:E1_286}).

Initially, this candidate was regarded as a possible 
$1.9$\,$R_{Earth}$-sized planet around a main sequence solar-like 
star. However, reconnaissance spectroscopy performed with SANDIFORD in 
February 2009 suggests that the target is a G8/9\,IV star. Additional 
observations with HIRES and HARPS indicate \teff\,$=5250\pm80\,K$, 
\logg\,$=3.75\pm0.10$, $[M/H]=-0.10\pm0.05$, and 
\vsini\,$=3.0\pm1.0$\,km\,s$^{-1}$, compatible with an evolved star 
of $M_*\approx1.1\,M_\odot$ and $R_*\approx2.3\,R_\odot$. This excluded 
a terrestrial-sized object but a transiting planet with 
$R_P\approx4.4\,R_{Earth}$ was still possible.

\begin{figure}
\resizebox{\hsize}{!}{\includegraphics{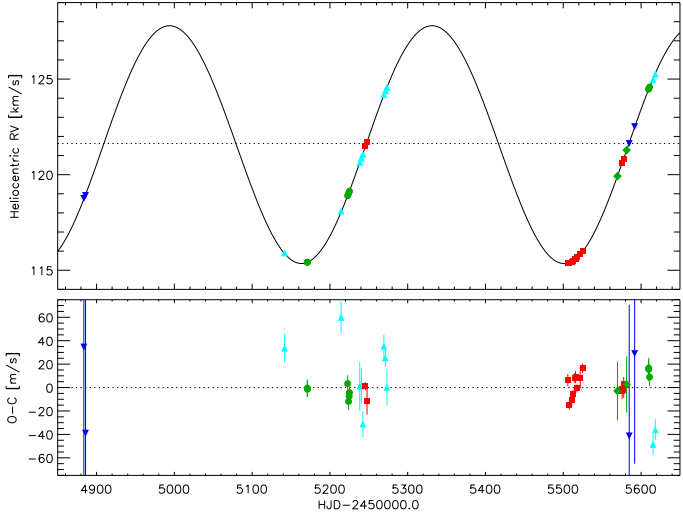}}
 \caption{\emph{Top panel}: RV curve of LRa01\,E1\,0286 as observed with Sandiford (blue downward 
          triangles), SOPHIE (light blue upward triangles), HIRES (green circles), HARPS (red 
	  squares), and FIES (green diamonds). The solid line represents the best fitting Keplerian 
	  orbital fit to the data (see text for more details). A RV shift was let free to vary 
	  in the fit between the five data-sets. The systemic RV of $V_\gamma=121.603$~\kms, 
	  as derived from the HARPS data-set only, is plotted with a horizontal dotted line. 
	  \emph{Bottom panel}: The RV residuals after subtracting the orbital solution (see the 
	  on-line edition of the Journal for a colour version of this figure).}
  \label{fig:RV_E1_286}
\end{figure}

RV observations carried-out over two years with SOPHIE, HIRES, HARPS, FIES, 
and SANDIFORD revealed a long-term RV trend indicative of an SB1 not in 
phase with the \corot\ transit ephemeris. The orbital fit to the data is 
shown in Fig.~\ref{fig:RV_E1_286}. The orbit seems nearly circular 
($e\approx0.01$) with a period of $P=337.52\pm0.20$ days and has a RV 
semi-amplitude $K=6.22\pm0.18$\,km\,s$^{-1}$. The derived mass function 
$f(m)\approx0.0084\,M_\odot$ implies a companion mass of approximately 
$0.22\,M_\odot$ assuming $1.1\,M_\odot$ for the primary star's mass.

\begin{figure}
\rotatebox{0}{
\resizebox{\hsize}{!}{\includegraphics{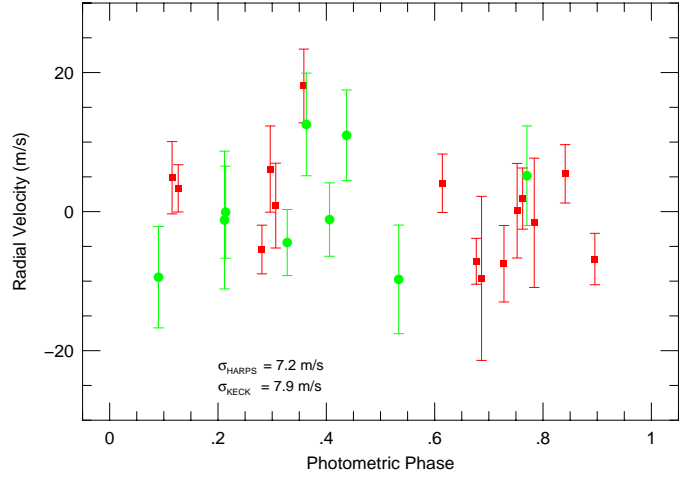}}
}
  \caption{The HARPS (red squares) and HIRES (green circles) RV residuals of LRa01\,E1\,0286 phased to the 
           \corot\ transit period P=3.60 days and epoch. The root main squares to the binary 
	   orbit fit, when considering only the HARPS and HIRES data, are reported in the 
	   labels (see the on-line edition of the Journal for a colour version of this figure).}
  \label{fig:RV_E1_286resids}
\end{figure}

The residuals to the binary orbit fit have a rms scatter of about 
7.5\,m\,s$^{-1}$ when using only the higher quality HARPS and HIRES data. 
Fig.~\ref{fig:RV_E1_286resids} shows the residual RV measurements phased 
to the \corot\ transit period. There is no convincing sinusoidal variations. 
The RV semi-amplitude of a possible planetary companion orbiting the target 
star would have to be less than about 5\,m\,s$^{-1}$. 

Recent time-series photometric observations performed with MegaCam@CFHT3.6m 
and EulerCam@Euler1.2m suggest that a faint, nearby star ($R\approx18$~mag, 
5\arcsec\ South from the target) might experience eclipses of $\sim5$\,\% 
that would account for the detected 0.03\,\% transit-like signal on LRa01\,E1\,0286. 
However, the photometric follow-up is still not conclusive and further 
investigation are needed to assess the real nature of the candidate. 
In conclusion, the target star is part of a long period binary and the 
transit signal remains unresolved. The LRa01\,E1\,0286 candidate might be 
either a planet in a stellar binary system or a contaminating eclipsing 
binary.

\subsubsection{LRa01 E1 2101 - CHR - 0102568803}
\label{LRa01 E1 2101}

According to Exo-Dat, this $V=14.15$\,mag target is of spectral 
type K1\,III. The \corot\ lightcurve shows a 2\,\% sun-spot-induced  
variability with a period of $\sim$11~days. LRa01\,E1\,2101 appears 
to be orbited by a transiting object ($D=0.08$\,\%, $P=2.72$~days). 
The signal is only significant in the red \corot\ photometric 
channel. As already described in Section~\ref{sec:Data}, due to the 
star's PSF on the CCD and the choice of the photometric mask, the 
expected depth of the transit in the green and blue \corot\ photometric 
channels is smaller than the scatter of the data-points, preventing 
any comparison of the transit depth in the different channels. 
Therefore, it cannot be concluded from \corot\ data alone that the 
candidate is a contaminating eclipsing binary. The fist HARPS spectrum 
reveals a narrow single peak CCF (FWHM=8\,\kms) and a K6\,V star with 
\teff$\approx$4250\,K and \logg$\approx$4.5\,dex 
($M_{*}$$\approx$$0.7\,M_{\odot}$, $R_{*}$$\approx$$0.7\,R_{\odot}$). 
The shallow transit depth thus implies a companion radius of 
about 2\,$R_{\oplus}$. But the transit is V-shaped suggesting a grazing 
transit/eclipse. Combined measurements with CFHT and MONET-North exclude 
background eclipsing binaries. The transit is considered to be on target. 
Six RV measurements acquired with HARPS show no significant sinusoidal 
variation down to a precision of 18~m\,s$^{-1}$. Follow-up is ongoing.

\subsubsection{LRa01 E1 2240 - CHR - 0102698887}

The transit signal occurs every 2.03 days in the lightcurve of 
this $V=15.22$\,mag target, classified as a F8 sub-giant according to 
Exo-Dat. The shallow (0.09\,\% deep) signal is a little asymmetric
when phase-folded. However, due to the low S/N ratio, the 
transit shape cannot be used to rule out a no-planet scenario. If the 
candidate is a planet, the transit depth suggests a Neptune-like 
planetary radius. No follow-up observations have been made yet.

\subsubsection{LRa01 E1 3216 - MON - 0102754163}

This is a faint $V=15.7$\,mag A5\,IV candidate (Exo-Dat) with a 
periodic 3.11 days transit signal of depth 0.13\,\%. It shows shallow 
out of transit variations. No ground-based follow-ups have been 
performed yet. 

\subsubsection{LRa01 E1 3221 - MON - 0102634864}

This is a rather long-period candidate with $P=32.33$~days. The transit 
is V-shaped and 2.33\,\% deep. This target of brightness $V=15.58$\,mag 
is listed in Exo-Dat as a A5 dwarf star. However, pulsations with 
periods of 0.78 and 8.75 days typical for giant stars have been 
detected in the \corot\ lightcurve, in disagreement with the main 
sequence scenario. No follow-up observations have been made.

\subsubsection{LRa01 E1 4423 - MON - 0102782651}

A faint ($V=16.22$\,mag) candidate showing a V-shaped shallow transit 
($D=0.25$\,\%) with a period of 1.87 days. No follow-up observations have 
been carried out. The \corot\ lightcurve is strongly affected by 
instrumental effects (jumps). Exo-Dat list the spectral type as K4\,V.

\subsubsection{LRa01 E1 4594 - MON - 0102617334}

This transiting candidate ($D=0.27$\,\%) was not 
spectroscopically observed due to the faintness of the target star 
($V=16.66$\,mag, SpT =O8\,III from Exo-Dat). IAC80 photometry excludes 
background eclipsing binaries. The transit duration of 6.6 hours for a 
transit period of $P=5.49$~days is consistent with an evolved 
host star.

\subsubsection{LRa01 E1 4667 - MON - 0102588881}

A 1.52\,\% deep transit signal with a period of 27.29 days is 
found in the lightcurve of this $V=16.08$\,mag star of spectral type 
A5\,IV (Exo-Dat). Wise photometric observations are inconclusive due 
to bad weather. IAC80 observations exclude contaminating eclipsing 
binaries. Two RV measurements acquired with HARPS at photometric phases 
0.43 and 0.76 shows a radial velocity variation of 84~m\,s$^{-1}$ which 
is comparable to the errors (i.e., $\sim$70\,\ms). The HARPS spectra 
unveils a G0\,V star, in disagreement with Exo-Dat. Assuming 
$M_*$=1\,$M_{\odot}$ for the host-star, a Jupiter-mass planet
in a 27.29 days orbit around LRa01\,E1\,4667 would produce a 
peak-to-peak RV variation of $\approx120$\,\ms which is almost twice 
the HARPS errors. Although the RVs exclude a stellar/brown dwarf 
companion to LRa01\,E1\,4667, considerably more HARPS measurements 
are required to assess the real nature of the transiting object. 
Follow-up is ongoing.

\subsubsection{LRa01 E1 4719 - MON - 0102703155}

A 1.26 days period transit signal of depth 0.10\,\% is found in
the \corot\ light curve of this $V=15.88$\,mag candidate (SpT=F8\,IV, 
Exo-Dat). The transit is V-shaped and asymmetric when phase-folded. 
However, the S/N ratio of the detected events is low. 
The shape may be distorted by either stellar activity or photometric 
noise. EulerCam could not observe the transit on target. This is 
expected given the shallow transit depth. Although the transit timing 
error at the time of the EulerCam observations (28 October 2010) was 
$\pm2$~hours and there is a risk that the transit might have been 
missed, it is concluded from photometric on-off observations that 
large variations by contaminants are probably not the cause for the 
transit event. No RV-measurements, however, have been acquired.

\subsubsection{LRa01 E1 4820 - MON - 0102751316}

This is a transiting candidate with a depth of 0.46\,\% and a 
period of 1.61 days. Photometric follow-up with the ESA-OGS facility 
confirms that the transit is on target. Analysis of the lightcurve 
indicates possible depth differences between odd and even transits 
and out-of-transit variation. The A5\,IV star (Exo-Dat) is too 
faint ($V=16.15$\,mag) for RV confirmation.

\subsubsection{LRa01 E1 5320 - MON - 0102666452}

This is another transiting candidate around a faint star 
($V=16.13$\,mag, SpT=G2\,V; Exo-Dat). The 0.14\,\% deep signal 
can only be identified when phase-folded with a 1.97~days period. 
The 3.3~hours transit duration appears to be a little long for a 
planetary object. No ground-based observations have been made.

\subsubsection{LRa01 E1 5536 - MON - 0102670085}

The \corot\ lightcurve of this $V=16.21$\,mag star of spectral type F8\,IV 
(Exo-Dat), shows a 0.27\,\% deep transit signal occurring every 0.90~days. 
CFHT photometric observations find a $0.40\pm0.25$\,\%  deep transit on 
target, compatible with the \corot\ signal. However some background stars, 
spatially located around the target, cannot be excluded as possible 
contaminants and the observations are considered inconclusive. No 
spectroscopic data have been acquired.

\subsubsection{LRa01 E2 3156 - CHR - 0102716818}
\label{LRa01 E2 3156}

The target ($V=15.76$\,mag) is a K2\,III star according to Exo-Dat. The 
0.15\,\% deep transit signal appears every 1.47 days and is detected only 
in the red \corot\ channel. As for LRa01\,E1\,2101 candidate 
(Section~\ref{LRa01 E1 2101}), also in this case the photo-noise in the 
green and blue channels is larger than the expected depth. The transit is 
V-shaped, and the duration of 2 hours is quite long for a planet. Observations 
with IAC80 exclude contamination by neighbouring objects. The transit is 
likely on target with a 20-30\,\% chance of a missed transit due to timing 
error. Two and seven RV measurements with HARPS and HIRES, respectively, 
show no significant variations down to a precision of 10~m\,s$^{-1}$. 
The candidate is still under investigation.

\subsubsection{LRa01 E2 3619 - MON - 0102765275}

V-shaped transit signals of 6\,\% depth with a period of 50.91 days are 
found in the lightcurve of this star ($V=15.56$\,mag). The spectral type is 
G8\,V, according to Exo-Dat. Low-resolution spectroscopy performed with 
AAOmega indicates that the target is a G0\,IV/V star. However the \corot\ 
photometric data show multi-periodic variations that seem more consistent 
with a giant. Therefore, the true spectral type is unclear. This candidate 
is also found in the IRa01 field \citep[as IRa01~E1~2060; ][]{Carpano2009}.

\subsubsection{LRa01 E2 4519 - MON - 0102580137}

A 0.14\,\% deep candidate with a period of 2.37 days is detected in 
the lightcurve of this A5\,IV star (Exo-Dat) candidate with magnitude 
$V=15.75$\,mag. EulerCam and IAC80 find no relevant photometric variations in 
any of the nearby stars. However, the transit may have been missed due to 
large (1.5~hours) timing errors. Still, the transit is considered to 
be likely on target. No RV measurements have been acquired for this star.

\subsection{Unsettled low priority planetary candidates}
\label{Sec: Unsettled low priority candidates}

Many likely binary candidates are discovered at the photometric level 
and have a low priority in the follow-up observation chain. Consequently, 
many of these were not observed by the follow-up team, especially when 
the target star is faint.

\subsubsection{LRa01 E1 2970 - CHR - 0102625386}

The $V=14.49$~mag candidate shows a V-shaped, 0.62\,\% deep transit 
signal occurring every 34.10 days. The transit is only seen in the \corot\ 
red channel. No event is detected in the blue and green channels at 
4 and 5\,$\sigma$ significance, respectively. This is a characteristic 
sign of a contaminant eclipsing binary. Hints of secondary eclipse are also
detected in the red lightcurve. Exo-Dat lists the spectral type as A5\,IV. 
No follow-up observations have been performed for this star as the candidate 
is suspected to be a contaminating eclipsing binary (CEB).

\subsubsection{LRa01 E1 3617 - MON - 0102617210}

According to Exo-Dat this is a A0\,V star ($V=15.62$\,mag). The 
0.64\,\% deep signal has a period of 2.73 days. Several coherent 
frequencies are found which hint to stellar variability induced by 
a massive companion. Hints of a secondary shallow eclipse have been 
recently found in the \corot\ lightcurve. No follow-up observations 
were carried out as a binary system is suspected for this candidate.

\subsubsection{LRa01 E1 3674  - CHR - 0102732757}

A V-shaped transit signal is found in the lightcurve of this 
A0\,V star ($V=15.32$\,mag, Exo-Dat). The transit duration of 
3.38 hours is quite long for a 1.97 days transit period. The 
0.20\,\% deep signal can only be seen in the red \corot\ 
channel. It is not detected, neither in the green nor in the
blue channel, with a $\sim$3\,$\sigma$ significance. No 
follow-up observations have been performed on this candidate
as a CEB scenario is suspected.

\subsubsection{LRa01 E1 4272 - MON - 0102626872}

The detected transit of this object is 2.44\,\% deep with a 1.88 
days period. In Exo-Dat the star is listed as A0\,V with a 
brightness of $V=15.87$\,mag. Photometry with ESA-OGS shows 
the transit to be on target. Due to out-of-transit variations in 
the lightcurve a stellar binary is suspected.

\subsubsection{LRa01 E1 4777 - CHR - 0102620061}

A V-shaped transit signal with a period of 3.35 days and a depth 
of 1.96\,\% is detected in the lightcurve of this $V=15.26$\,mag 
candidate (A5\,IV, Exo-Dat). The duration (3.90 hours) is rather 
long for a V-shaped transit and the depths observed in the three 
\corot\ colors differ by more than $1\,\sigma$. IAC80 observations 
performed in February 2011 shows no variations neither on target 
nor on any of the nearby stars. An underestimate of the transit 
timing error, listed at about 30 minutes at the time of the IAC80 
observations, might account for the no ground-based transit 
detection. No RV follow-up observations have been carried out for 
this candidate.

\subsubsection{LRa01 E1 4836 - MON - 0102630623}

Analysis of this V-shaped transit candidate ($D=4.70$\,\%, $P=36.78$ 
days) shows significant depth difference between even and odd 
transits ($12 \sigma$). It is therefore a low priority candidate 
for which no follow-up observations have been carried out. Exo-Dat 
lists the spectral type of this $V=15.85$\,mag star as A5\,V.

\subsubsection{LRa01 E1 5450 - MON - 0102595916}

The 0.22\,\% deep transit signal in the lightcurve of a G0\,IV 
star of brightness $V=16.38$\,mag (Exo-Dat) appears to have an 
asymmetric shape. In addition, the transit duration of 9.23 hours 
is too long for the 4.11 days transit period to be consistent with 
a planet. Due to the faintness of the star and the bad transit 
properties the candidate has a low priority in the follow-up chain 
and was not observed.

\subsubsection{LRa01 E2 2185 - MON - 0102729260}

A G2\,V star with $V=15.08$\,mag (Exo-Dat) shows V-shaped transit 
signals with a depth of 0.24\,\% and a period of 1.69 days. The
candidate is already known from the \corot\ IRa01 run as 
IRa01\,E1\,1319 \citep{Carpano2009}. The transit duration of 3.57 
hours seems to be quite long for a planetary object. In addition, 
a secondary eclipse at phase=0.5 and depth differences between 
even and odd transits were detected. This is most likely a binary 
system. Follow-up observations for this candidate are not foreseen.

\subsubsection{LRa01 E2 2597 - CHR - 0102672065}

A 8.90 days transit signal is found in the lightcurve of 
this $V = 14.17$\,mag star. AAOmega spectroscopy shows the 
candidate to be a G6\,III/IV star, in good agreement with 
the Exo-Dat classification (G5\,III). FLAMES yields 
\teff\,$=4991\pm140$~K, \logg\,$=3.24\pm0.30$~dex, 
$[m/H]=-0.29\pm0.15$~dex, and \vsini\,$=4.8\pm2.0$~km\,s$^{-1}$ 
\citep{Gazzano2010}. As described in Section~\ref{sec:Data}, 
the deep signal (1.00\,\% when normalized to the blue flux only) 
is seen only in the \corot\ blue channel (Figure~\ref{fig:BEB}). 
The event is detected, neither in the green nor in the red 
lightcurve, with a 12\,$\sigma$ and 25\,$\sigma$ significance, 
respectively. This indicates that the candidate is with high 
probability a contaminating eclipsing binary (CEB).

\subsubsection{LRa01 E2 2627 - CHR - 0102757559}

According to low-resolution spectroscopy performed with AAOmega 
the spectral type of this $V=15.13$\,mag star is F4\,V, whereas 
Exo-Dat lists this target as a G2\,V object. A V-shaped deep signal 
(0.083\,\%) is found only in the \corot\ blue channel, with a 
significant not-detection in the green and red lightcurves 
(4\,$\sigma$). This indicates a contaminating eclipsing binary with 
a period of 0.95 days. Neither photometric nor further spectroscopic 
follow-up is foreseen.

\subsubsection{LRa01 E2 3157 - CHR - 0102672700}

The object is a low priority 0.24\,\% deep candidate ($P=1.87$~days) 
due to 1) V-shaped transit curve, 2) differences in the transit depths at
a $5\,\sigma$ significance, and 3) the transit is only seen in the red 
channel with a $\sim$3\,$\sigma$ significant no-detection in the blue and 
green colors, suggesting the presence a nearby contaminating eclipsing 
binary. Exo-Dat lists the spectral type of this $V=14.86$\,mag star as 
G0\,IV. No follow-up observations have been carried out for this star as
a CEB scenario is suspected.

\subsubsection{LRa01 E2 4494 - MON - 0102587927}

A shallow transit (0.13\,\%) signal with a period of 
about 2~days) is found in the \corot\ lightcurve of this faint 
($V=16.07$\,mag) target. Exo-Dat list the spectral type as 
K3\,V. The apparent transit shape is asymmetric and the
transit duration (2.81 hours) is quite long for the
orbital period. No follow-up observations have been 
performed on this candidate.

\subsubsection{LRa01 E2 4910 - MON - 0102780627}

This is a candidate around a star of brightness 
$V=15.36$\,mag listed as a F8 dwarf in Exo-Dat. It has
been also detected in the IRa01 run (IRa01\,E1\,1531) and listed as a
planetary candidate in \citet{Carpano2009}. Observations 
with AAOmega classify this candidate as a F7/8 dwarf star, 
in really good agreement with Exo-Dat. The transit is 
0.87\,\% deep and appears every 2.38 days. IAC80 observations 
confirms the transit signal is on target. Due to secondary 
faint eclipses at phase 0.5 recently detected in the lightcurve, 
it is suspected to be a binary. No further follow-up is foreseen.

\subsubsection{LRa01 E2 5194 - MON - 0102604000}

This candidate is suspected to be a binary. The 2.8 hours 
duration of the 0.68\,\% deep transit is too long for a transiting 
planet with an orbit period of 1.25 days around a K2V star (Exo-Dat).
No spectroscopic follow-up observations have been made due to 
the faintness of the star ($V=16.09$\,mag) and the lack of 
ground-based confirmation that the transit is on-target.

\subsection{False alarms}
\label{Sec: False alarms}

Sometimes, \corot\ lightcurves are affected by instrumental effects 
(e.g., jumps, glitches) making the transit detection uncertain.

\subsubsection{LRa01 E1 2960 - CHR - 0102613782}

There appears to be a 0.17\,\% deep transit with a 13.03 days period 
in this lightcurve of a $V=14.40$\,mag F8\,IV star (Exo-Dat).
This signal can only be found in the \corot\ red channel which is 
affected by instrumental effects (jumps). If real and on target 
the transit should be visible in the green and blue channel as 
well. Therefore, it is suspected to be a false alarm.

\subsubsection{LRa01 E2 3389 - CHR - 0102674894}

This shallow transit signal ($D=0.08$\,\%, $P=7.03$~days) is found in 
the lightcurve of a faint ($V=15.65$\,mag) G0\,III star (Exo-Dat), 
but the lightcurve is disturbed by instrumental effects (jumps). 
It is suspected to be a false alarm.

\subsubsection{LRa01 E2 3612 - MON - 0102577194}

A 0.34\,\% deep transit signal with a period of 38.24 days is found 
in the lightcurve of this faint object ($V=16.01$\,mag), listed as 
a G0\,IV star in Exo-Dat. The shape of the transit is asymmetric and 
the data suffer from glitches and jumps. This is likely a false
alarm.

\subsection{X-cases}
\label{Sec: X-cases}

The following stars are objects that might be planetary candidates 
if the spectral type of the target was considerable different than 
listed in Exo-Dat.

\subsubsection{LRa01 E2 0928  - MON - 0102664130}

The target star of this 2.60\,\% deep transit candidate is according 
to Exo-Dat a A5\,IV star, implying a stellar radius of about 
2~$R_\odot$ \citep{Cox2000}. If true, the V-shaped signal, occurring 
every 49.9 days, cannot be caused by a planet. But the true spectral 
type of a star can differ significantly from the one given in Exo-Dat. 
Therefore the candidate was not completely discarded as a binary 
pending the confirmation of the spectral type. Unfortunately, the 
star is very faint ($V=15.62$\,mag) making a rapid spectral type 
classification difficult.

\subsubsection{LRa01 E2 5678 - MON - 0102613411}

Another deep candidate which might be a planetary candidate if the 
true stellar radius is smaller than listed in Exo-Dat (i.e., 
$R_*\approx1.1\,R_\odot$ for a F8\,V star). The transit is 
4.33\,\% deep and occurs every 18.76 days. Unfortunately, the 
star is relatively faint ($V=15.91$\,mag) making it a difficult 
target for a rapid spectral type classification.

\subsection{Mono-transits}
\label{Sec: Mono-transits}

\subsubsection{LRa01 E1 2765 - MON - 0102647266}

A mono-transit candidate is found with $D=8.30$\,\% at epoch 
$HJD=2454465.434$ in the light curve of a A5\,V star of 
magnitude $V=14.52$\,mag (Exo-Dat).

\subsubsection{LRa01 E1 4785 - CHR - 0102709133}

This single transit ($D=9.70$\,\% at $HJD=2454419.798$) is 
more prominent in the blue channel and there might be a 
secondary smaller eclipse at $HJD=2454524.348$ with a 
depth of $\sim1$\,\%. This candidate is probably a
contaminating eclipsing binary with a period of 
$\sim209$~days. The target star is listed in Exo-Dat 
as a K2\,V star with magnitude $V=15.50$\,mag.

\subsubsection{LRa01 E2 1113 - CHR - 0102574444}

At epoch $HJD=2454422.342$ a 5.40\,\% deep transit can be 
found in the lightcurve of this star. The transit signal  
is deeper in the blue channel. Probably the transiting 
object does not orbit this F8\,IV star of brightness 
$V=14.14$\,mag (Exo-Dat). It is a suspected 
contaminating eclipsing binary.

\subsubsection{LRa01 E2 2368 - MON - 0102582649}

A single very deep transit/eclipse ($D=17.80$\,\%) is found 
at epoch 2454510.992 HJD. The $V=14.98$\,mag object is 
listed in Exo-Dat as F6\,V star.

\subsubsection{LRa01 E2 2744 - MON - 0102586624}

At epoch $HJD=2454473.174$ a 15.75\,\% deep single transit/eclipse 
is visible in the lightcurve of this object. The star is listed 
in Exo-Dat as A5\,IV and magnitude $V=15.11$\,mag.

\section{Discussion}
\label{sec: discussion}

\begin{figure}[t]
\begin{center}
    \includegraphics[width=1.00\linewidth]{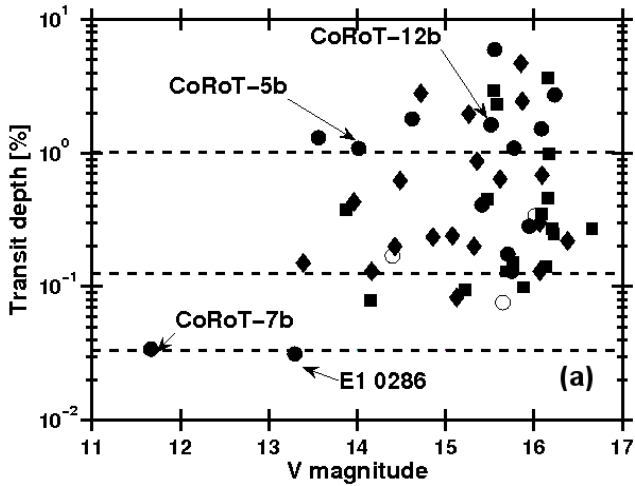}
  \caption{Transit depth ($D$) versus $V$ magnitude of the candidates found
           in LRa01 by the detection group after exploiting the full length of
	   the lightcurves. Filled circles are good planetary candidates, squares are 
	   low priority candidates, diamonds are suspected binaries, and open circles 
	   are suspected false alarms. Arrows mark the three transiting planets found 
	   in the LRa01 run: \corot-5b, \corot-7b and \corot-12b, and the suspected 
	   planet LRa01\,E1\,0286. The horizontal dashed lines represent (from top to bottom) 
	   the expected signal produced by a Jupiter-size planet, a Neptune-size planet 
	   and a 2 Earth radii planet around a solar-like star, respectively.}
  \label{fig:depth_MV}
\end{center}  
\end{figure}

\begin{figure}[t]
\centering
 \includegraphics[width=0.9\linewidth]{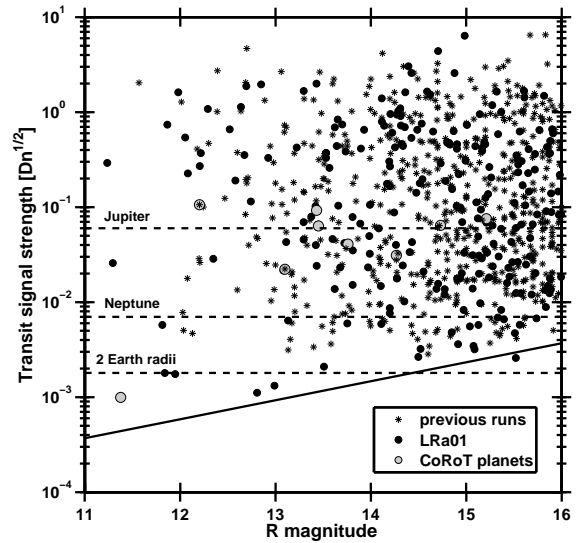}
  \caption{Transit signal strength versus $r^\prime$ magnitude of detected planetary and 
           binary candidates for the previous IRa01, SRc01, LRc01 \corot\ runs (asterisks), 
	   LRa01 run (filled circles), and first seven \corot-planets (open circles). $D$ 
	   is the transit depth and $n$ is the number of points in a transit event. The 
	   horizontal dashed lines represent (from top to bottom) the expected signal produced 
	   by a Jupiter-size planet, a Neptune-size planet, and a 2-Earth radii planet around
	   a solar-like star, respectively. Transit candidates and binaries have been combined 
	   in this example to have a better statistical basis of the detection efficiency. The 
	   solid line represents the photon noise.}
  \label{fig: rednoise}
\end{figure}

\begin{figure}[ht]
\centering
 \includegraphics[width=0.9\linewidth]{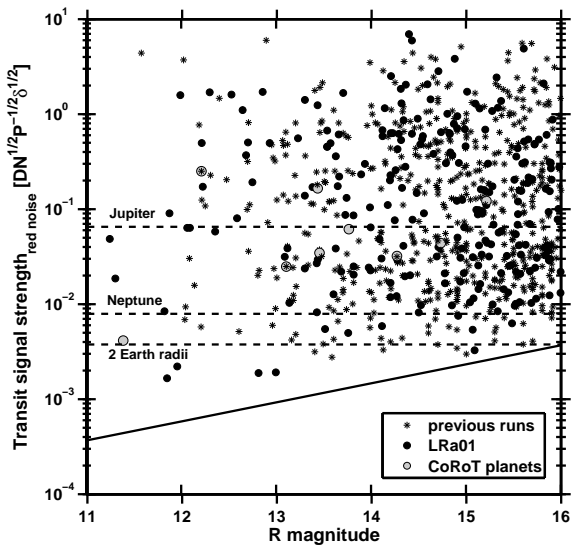}
  \caption{Transit signal strength in the presence of red noise versus $r^\prime$ magnitude for 
           the detected and binaries candidates in the previous IRa01, SRc01, LRc01 \corot\ runs 
	   (asterisks), LRa01 run (filled circles), and the first seven \corot-planets (open circles). 
	   $D$ is the transit depth, $N$ is the number of data points in a lightcurve, $P$ is the transit 
	   period, and $\delta=512$~seconds is the sampling interval. The horizontal dashed lines represent 
	   (from top to bottom) the expected signal produced by a Jupiter-size planet with 3 days orbit period, 
	   a Neptune-size planet with 3 days orbit period and a 2 Earth radii planet with 1 day orbit period
	   around a solar-like star, respectively. The solid line represents the photon noise.}
  \label{fig: rednoise_2}
\end{figure}

\corot-7b \citep{Leger2009} and the candidate LRa01\,E1\,0286 
proves the capability of \corot\ for discovering Super-Earths 
around main-sequence stars. Although the latter candidate is 
maybe not a Super-Earth, the depth of the detected transit is 
comparable to the signal expected from a \emph{bona fide} 
Super-Earth around a solar-like star. Therefore, it is 
concluded from Figure~\ref{fig:depth_MV} that \corot\ is 
capable of detecting Super-Earths with periods in the range 
of one to four days around stars of apparent magnitude 
$V\leq13.3$\,mag. However, most of the target stars are 
fainter than this limit (Figure \ref{fig:m_V}). The 
\corot-7b case was favourable for the terrestrial-size planet 
detection. Indeed the planet orbits a star of magnitude 
$V=11.65$\,mag in less than one day. \corot-7 is one of 
the brightest objects in the LRa01 star field. The follow-up 
for the candidate LRa01\,E1\,0286 proved to be much more 
difficult although the star of magnitude $V=13.30$\,mag 
is still quite bright compared to others in the field.

Compared to previous \corot\ runs (i.e., IRa01, SRc01, 
and LRc01) more and weaker candidates have been found in 
the LRa01 field (Figure~\ref{fig: rednoise} and 
\ref{fig: rednoise_2}). This is to be expected, since 
LRa01 covers a longer time period than SRc01 and IRa01 
($\approx 25$ days for SRc01 and $\approx 90$ days for 
IRa01). The lack of small transit candidates around bright 
stars in the LRc01 star field is explained by the unfavourable 
stellar population properties of this particular star field 
for the search of extrasolar planets: 58\,\% of all stars 
were identified as giants. Furthermore, LRc01 contains in 
total less stars with apparent magnitude $V<13$\,mag. On 
the other hand, the LRa01 star field contains at least 
62\,\% main sequence stars. Overall, it contains more 
bright main sequence stars with apparent magnitude 
$V<13$\,mag than previous long runs (Section~\ref{sec: field} 
and Table~\ref{tab: bright_stars}). Planets around such 
stars are easier to detect in the photometric data and 
easier to observe by ground-based telescopes. From
the perspective of the types of false positives and 
\emph{bona fide} transiting planets identified in LRa01 
with respect to IRa01 and LRc01, we found that the rates 
of discovered planets, spectroscopically confirmed eclipsing 
binary systems, CEB, and blends is comparable between the 
three runs.

The lack of confirmed Neptune-size planets is probably 
not due to limitations in the \corot\ follow-up or detection 
chain. Although the follow-up is constrained by the 
limited observation time as pointed out by \citet{Moutou2009}, 
\corot\ was able to find at least one Super-Earth and 
other planet search programs yield similar result. 
Short-period Neptune-size planets may be very rare 
objects, as tentatively pointed out 2005 by 
\citet{Mazeh2005} and as seems to be confirmed by the 
distribution of Kepler planetary candidates 
\citep{Borucki2011}. Such planets may evaporate faster 
than more massive gas giants because their low surface 
gravity is not able to maintain the irradiated 
atmosphere \citep{Southworth2007}. \citet{Davis2009} 
provide further evidence for a lost population of 
short-period Neptune planets. \citet{Almenara2009}, after 
analysis of three \corot\ fields, further support the 
conclusion that the lack of confirmed Neptune planets 
in the \corot\ fields is not due to observational 
limitations.

\begin{table}[t]
\centering
\caption{Number of bright main sequence stars in LRa01 and the previous runs.}
\label{tab: bright_stars}
\begin{tabular}{cc}
\hline
\hline
\noalign{\smallskip}
\corot\ run & Number of dwarf stars  \\
            & with $V<13$\,mag       \\
\hline
\noalign{\smallskip}
IRa01 & 308\\
LRc01 & 138\\
LRa01 & 438\\
\noalign{\smallskip}
\hline
\end{tabular}
\end{table}

\section{Summary}
\label{sec: summary}

\corot\ observed the LRa01 star field continuously for 130 days 
and collected 11\,408 lightcurves. There are 7\,470 chromatic 
photometric data-sets and 3\,938 monochromatic data-sets. The 
\corot\ detection group performed a full in-depth analysis of 
the lightcurves. 242 lightcurves (2.1\,\% of all lightcurves) 
contain a transit signal. 191 of these (79\,\% of the candidates or 
1.7\,\% of all stars in the field) were identified as binaries 
based on photometric analysis only (Table~\ref{Tab:Binaries}), 
including five mono-transits (Section~\ref{Sec: Mono-transits}).

Fifty-one signals were classified as planetary candidates and 
proposed for observational follow-up with different priorities 
based on the photometric analysis. Thus in about 0.5\,\% of all 
\corot\ targets a signal was detected that might originate from 
a planetary transit. In addition, 3 candidates were discarded 
as likely false alarms based on photometric analysis 
(Section~\ref{Sec: False alarms}). Five mono-transits were 
detected with depths compatible with an eclipsing binary 
(Section~\ref{Sec: Mono-transits} and Table~\ref{Tab:Binaries}). 
Two candidates were classified as potential planetary candidates 
or ``X-cases'' (Section~\ref{Sec: X-cases}), if the stellar radius 
of the target star is significantly smaller than listed in the 
Exo-Dat database. None of the false candidates and X-cases were 
followed-up and are included for completeness. See also 
Table~\ref{Tab:all_planets}.

Of the fifty-one candidates, thirty-seven (73\,\% of all candidates) 
are ``good'' planetary candidates based on photometric analysis only 
(Sections~\ref{Sec:confirmed planets}, \ref{Sec:Settled cases}, and 
\ref{Sec: Unsettled good candidates}). Thirty-two of the ``good'' 
candidates have been followed-up and the nature of twenty-two objects 
has been solved. Four candidates (about 8\,\% of all candidates) have 
been confirmed as transiting planets (Table \ref{tab: planetary_param}): 
\corot-7b \citep{Leger2009}, \corot-5b \citep{Rauer2009}, \corot-12b 
\citep{Gillon2010}, and the recently confirmed hot-Jupiter LRa01\,E2\,5277 
\citep[\corot-21b,][]{Patzold2011}. Another two non-transiting planets were 
detected by RV measurements only: \corot-7c \citep{Queloz2009} and 
\corot-7d \citep{Hatzes2010}. Another candidate, LRa01\,E1\,0286 may be 
a planetary object in a binary system but is unconfirmed yet. Eighteen 
objects (49\,\% of the good candidates) were identified as non-planetary 
objects. Six are contaminating eclipsing binaries (CEBs) and two are 
blends (i.e., LRa01\,E1,1123 and LRa01\,E2\,5184). Six candidates 
could be resolved spectroscopically as stellar binaries (SB). Four 
candidates are stellar companions around early-type stars. 
 
According to the lightcurve analysis only, fourteen candidates (27\,\% 
of all candidates; Section~\ref{Sec: Unsettled low priority candidates}) 
have low priorities because of one or more characteristics hinting at a 
non-planetary scenario: out of transit variations, depth differences 
between even and odd transits, depth differences in different color 
channels, and very shallow secondary eclipse. Four of these were 
followed-up but the observations are not conclusive. 

The follow-up for most of the LRa01 candidates is now concluded. Only 
LRa01\,E1\,0286, LRa01\,E1\,2101, LRa01\,E1\,4667, and LRa01\,E2\,3156 are 
still under investigation.

\begin{acknowledgements}

We thank the anonymous referee for her/his careful reading, useful comments, 
and suggestions, which helped to improve the manuscript.

The German \corot\ Team (University of Cologne and TLS) acknowledges 
\emph{Deut\-sches Zentrum f\"ur Luft- und Raumfahrt} 
(DLR) grants 50\,QM\,1004, 50\,OW\,0204, 50\,OW\,0603, and 50\,QP\,07011. 
The team at the IAC acknowledges support by grants ESP2007-65480-C02-02 
and AYA2010-20982-C02-02 of the Spanish Ministerio de Ciencia e Innovaci\'on.

Partly based on observations carried-out at the European Southern 
Observatory (ESO), La Silla and Paranal (Chile), under observing programs 
numbers 080.D-0151, 081.C-0388, 081.C-0413, 083.C-0186, and 282.C-5015. 
The authors are grateful to the staff at ESO La Silla and ESO Paranal 
Observatories for their support and contribution to the success of the 
HARPS, UVES, NACO, and CRIRES observing runs.

Parts of the data presented herein were also obtained at the W.M. Keck 
Observatory from telescope time allocated to the \emph{National Aeronautics 
and Space Administration} through the agency's scientific partnership with 
the California Institute of Technology and the University of California. 
The Observatory was made possible by the generous financial support of the 
W.M. Keck Foundation. The authors wish to recognize and acknowledge the 
very significant cultural role and reverence that the summit of Mauna Kea 
has always had within the indigenous Hawaiian community. We are most 
fortunate to have the opportunity to conduct observations from this mountain.

The present manuscript is also based on observations performed with 
\emph{a)} the IAC80 telescope operated by the Instituto de Astrof\'\i sica 
de Tenerife at the Observatorio del Teide. We thank its observing staff; 
\emph{b)} the SOPHIE spectrograph at the Observatoire de Haute-Provence, 
France, under observing programs PNP.08A.MOUT, PNP.09A.MOUT, and PNP.10A.MOUT; 
\emph{c)} the FIES spectrograph at the Nordic Optical Telescope (observing 
program P42-261), operated on the island of La Palma jointly by Denmark, 
Finland, Iceland, Norway, and Sweden, in the Spanish Observatorio del Roque 
de los Muchachos of the Instituto de Astrofisica de Canarias; \emph{d)} the 
AAOmega multi-object facilities at the Anglo-Australian Telescope operated 
at Siding Spring Observatory by the Anglo-Australian Observatory (observing 
programs 07B/040 and 08B/003); \emph{e)} the SANDIFORD spectrograph at the 
2.1\,m Otto Struve telescope at McDonald Observatory of the University of 
Texas at Austin; \emph{f}) the MOnitoring NEtwork of Telescopes (MONET), 
funded by the ``Astronomie \& Interne'' program of the Alfried Krupp von 
Bohlen und Halbach Foundation, Essen, and operated by the 
Georg-August-Universit\"at G\"ottingen, the McDonald Observatory of the 
University of Texas at Austin, and the South African Astronomical 
Observatory.

The CoRoT/Exoplanet catalogue (Exo-Dat) was made possible by observations 
collected for years at the Isaac Newton Telescope (INT), operated on the 
island of La Palma by the Isaac Newton group in the Spanish Observatorio 
del Roque de Los Muchachos of the Instituto de Astrof\'\i sica de Canarias.

\end{acknowledgements}

\bibliographystyle{aa} 
\bibliography{Detektion} 

\longtab{4}{
\begin{longtable}{crrc}
\caption{\label{rvtable} RV measurements for LRa01 planetary candidates.}\\
\hline
\hline
\noalign{\smallskip}
    BJD         &     RV~~~~    & $\sigma$\,RV~~~   & Instrument \\
   (days]       &   [\kms]  & [\kms]  &            \\
\noalign{\smallskip}
\hline
\endfirsthead
\caption{Continued.}\\
\hline
\hline
\noalign{\smallskip}
    BJD         &     RV~~~~    & $\sigma$\,RV~~~   & Instrument \\
    (days]      &   [\kms]  & [\kms]  &            \\
\noalign{\smallskip}
\hline
\noalign{\smallskip}
\endhead
\hline
\endfoot
\noalign{\smallskip}
\multicolumn{1}{l}{LRa01\,E1\,0544} \\
2454766.69555 & 16.879   & 0.038  & SOPHIE \\  
2454767.64784 & 16.780   & 0.056  & SOPHIE \\
\hline
\noalign{\smallskip}
\multicolumn{1}{l}{LRa01\,E1\,0561} \\
2454515.34284 & 43.753   & 0.161  & SOPHIE \\
2454526.33924 & -8.432   & 0.132  & SOPHIE \\
\hline
\noalign{\smallskip}
\multicolumn{1}{l}{LRa01\,E1\,5015} \\
2455162.72198 &  62.831 &  0.246 & HARPS \\
\hline
\end{longtable}
}

\longtab{5}{
\begin{center}
\begin{tiny}
\begin{longtable}{ccrrcrcr}
\caption{\label{Tab:all_planets} The \corot\ LRa01 planetary candidates: coordinates and transit parameters. The listed depth is relative to the combined total flux.}\\
\hline \hline
\noalign{\smallskip}
\corot\ ID & Win ID  &$\alpha$\,(J2000)&$\delta$\,(J2000)&        Epoch (HJD)         &   Period~~~~~~~~~~~   & Depth & Duration   \\
           &         &    [deg]~~~     &       [deg]~~~  &           [days]           &   [Days]~~~~~~~~~~~   & \%    & [Hours]    \\
\noalign{\smallskip}
\hline
\noalign{\smallskip}
\multicolumn{8}{c}{Confirmed planets}\\
\noalign{\smallskip}
\hline
\noalign{\smallskip}
 102764809 & E1 1031 &    101.27723    &      0.81527    & $2454400.19885\pm0.00020$  & $4.037896\pm0.000002$ & 1.461 &  2.81~~~ \\ 
 102708694 & E2 0165 &    100.95606    &     -1.06303    & $2454398.07670\pm0.00150$  & $0.853585\pm0.000024$ & 0.034 &  1.25~~~ \\ 
 102671819 & E2 3459 &    100.76568    &     -1.29645    & $2454398.62707\pm0.00036$  & $2.828042\pm0.000013$ & 1.744 &  2.57~~~ \\ 
 102725122 & E2 5277 &    101.05245    &     -0.29913    & $2454399.02987\pm0.00090$  & $2.724740\pm0.000140$ & 0.449 &  4.76~~~ \\
\hline
\end{longtable}
\end{tiny}
\end{center}
}

\longtab{6}{
\begin{tiny}
\begin{longtable}{ccccccp{1.9cm}p{1.9cm}p{5.65cm}}
\caption{\label{Tab:all_planets_cont} The \corot\ LRa01 planetary candidates: follow-up results.}\\

\hline 
\hline
\noalign{\smallskip}
 Win ID &   $B$  &   $V$  & $r^\prime$ & $i^\prime$ & Candidate  &        \multicolumn{2}{l}{~~~~~~~Follow-up facilities}                            & Comments \\
        &  [mag] &  [mag] &   [mag]    &    [mag]   & Nature$^a$ &           Photometry                & Spectroscopy                         &          \\
\noalign{\smallskip}
\hline
\noalign{\smallskip}
\endfirsthead

\caption{Continued.}\\
\hline 
\hline
\noalign{\smallskip}
 Win ID &   $B$  &   $V$  & $r^\prime$ & $i^\prime$ & Candidate  &        \multicolumn{2}{l}{~~~~~~~Follow-up facilities}                            & Comments \\
        &  [mag] &  [mag] &   [mag]    &    [mag]   & Nature$^a$ &           Photometry                & Spectroscopy                         &          \\
\noalign{\smallskip}
\hline
\noalign{\smallskip}
\endhead

\noalign{\smallskip}
\hline
\endfoot

\noalign{\smallskip}
\hline
\endlastfoot  

\noalign{\smallskip}
\multicolumn{9}{c}{\large{Confirmed planets}}\\
\noalign{\smallskip}
\hline
\noalign{\smallskip}
E1 1031 & 14.656 & 14.018 &   13.760   &   13.405   & \corot-5b   & IAC80                              & AAOmega, SOPHIE, HARPS               & \citet{Rauer2009}\\
E2 0165 & 12.524 & 11.668 &   11.378   &   10.924   & \corot-7b   & CFHT, CST/FASTCAM, VLT/NACO, IAC80 & AAOmega, FLAMES, CRIRES, HARPS, UVES & \citet{Leger2009,Queloz2009,Hatzes2010,Bruntt2010,Pont2011,Hatzes2011}\\
E2 3459 & 16.343 & 15.515 &   15.211   &   14.685   & \corot-12b  & IAC80                              & HARPS, HIRES                         & \citet{Gillon2010}\\
E2 5277 & 16.946 & 16.090 &   15.726   &   15.175   & \corot-21b  & IAC80                              & HARPS, HIRES                         & IAC80 confirms the transit is on target. F8 sub-giant (\teff\,$\approx$\,6100\,K, \logg\,$\approx$\,3.5). HARPS and HIRES RVs data confirm a hot-Jupiter planet with $M_{p}$$\approx$$2\,M_{Jup}$ \citep{Patzold2011}.\\
\noalign{\smallskip}
\hline
\noalign{\smallskip}
\noalign{\smallskip}
\multicolumn{9}{c}{\large{Settled cases: non planetary objects}}\\
\noalign{\smallskip}
\hline
\noalign{\smallskip}
E1 0544 & 14.021 & 13.392 &   13.132   &   12.750   & CEB         & EulerCam, IAC80                   & AAOmega, SOPHIE              	      & Classified as F7\,V by AAOmega. The star is a fast rotator according to SOPHIE measurements and show no RV variation down to a precision of 50~\ms. EulerCam and IAC80 photometry reveals that a $\sim4$ magnitude fainter star located 9\arcsec~West of the main target contaminates the lightcurve with deep eclipses ($D\approx20$\,\%).\\
\end{longtable}
${}^a$ The following abbreviations are used.\newline
CEB: Contaminating eclipsing binaries\newline
SB1: binary system with one component spectroscopically resolved\newline
SB2: binary system with both components spectroscopically resolved\newline
SB3: triple system with components spectroscopically resolved\newline

\end{tiny}
}

\longtabL{7}
{
\begin{landscape}
\begin{center}
\begin{longtable}{ccrrrrrccrr}
\caption{\label{Tab:Binaries} \corot\ eclipsing binary candidates found in LRa01.}
\\
\hline
\hline
\noalign{\smallskip}
\corot\ ID& Win ID  &$\alpha$\,(J2000)&$\delta$\,(J2000)&$V$~~~~~&  Period~~ &$\sigma$~Period &   Epoch [days]  & $\sigma$~Epoch & Duration  &     Depth      \\
          &         &    [deg]~~~     &       [deg]~~~  & [mag]\,&  [days]~~~&     [days]~~   &  (HJD-2454000)  &     [days]     & [hours]~  &      [\%]~     \\
\noalign{\smallskip}
\hline
\noalign{\smallskip}
\endfirsthead

\caption{Continued.}\\
\hline
\hline
\noalign{\smallskip}
\corot\ ID& Win ID  &$\alpha$\,(J2000)&$\delta$\,(J2000)&$V$~~~~~&  Period~~ &$\sigma$~Period &   Epoch [days]  & $\sigma$~Epoch & Duration  &     Depth      \\
          &         &    [deg]~~~     &       [deg]~~~  & [mag]\,&  [days]~~~&     [days]~~   &  (HJD-2454000)  &     [days]     & [hours]~  &      [\%]~     \\
\noalign{\smallskip}
\hline
\noalign{\smallskip}
\endhead

\hline
\endfoot

\hline
\endlastfoot  

102642448 & E2 3931 &	 100.61002    &     -1.03843    & 14.854 &  0.946162 &    0.000005    &    399.018180	&    0.000419	 &   2.20    &        5.0     \\
102671362 & E2 3276 &	 100.76310    &     -0.34620    & 15.369 &  0.861199 &    0.000014    &    398.875793	&    0.001268	 &   3.90    &       14.1     \\
102648472 & E2 1645 &	 100.64257    &     -0.57276    & 14.491 &  1.857077 &    0.000007    &    399.542170	&    0.000290	 &   3.54    &       22.0     \\
102645479 & E1 2138 &	 100.62651    &      0.96876    & 13.850 &  3.187703 &    0.000014    &    400.646979	&    0.000333	 &   3.90    &        7.1     \\
102681270 & E2 1693 &	 100.81449    &     -0.67801    & 14.674 &  2.986371 &    0.000012    &    400.722071	&    0.000354	 &   3.90    &       30.9     \\
102715227 & E1 1535 &	 100.99357    &     -0.17365    & 14.458 &  5.707336 &    0.000086    &    403.458115	&    0.001121	 &   4.30    &        1.6     \\
102669801 & E1 0709 &	 100.75499    &      0.24078    & 12.344 & 65.353042 &    0.001927    &    460.010802	&    0.001759	 &   3.60    &        4.5     \\
102648907 & E2 3238 &	 100.64484    &     -1.07534    & 14.526 & 18.778638 &    0.000225    &    412.732898	&    0.000695	 &   2.50    &        4.2     \\
\end{longtable}
\end{center}
\end{landscape}
}

\end{document}